\documentstyle[11pt,epsfig]{article} 
\def\nN{n_{\rm n}} \def\nC{n_{\rm c}} \def\nB{n_{\rm b}}  
\def\uC{u_{\rm c}} \def\uN{u_{\rm n}} 
\def\xiC{\zeta} \def\xiN{\xi} 
\def\muC{\mu^{\rm c}} \def\muN{\mu^{\rm n}} 
\def\rhoN{\rho_{\rm n}} \def\rhoC{\rho_{\rm c}} 
\def\Lamb{\Lambda} 
\def\wN{w^{\rm n}} \def\CC{N}  
\def\gamC{\gamma_{\rm c}}   \def\TE{T}  \def\fE{f^{\rm ex}} 
\def\fN{f^{\rm n}} \def\fC{f^{\rm c}} 
\def\fL{f^{\rm me}} \def\fD{f^{\rm ch}}

\def\sqr#1#2{{\vcenter{\hrule height.4pt\hbox{\vrule width.8pt height#2pt 
\kern#1pt\vrule width.8pt}\hrule height.4pt}}} 
\def\Square{\mathchoice{\sqr78\,}{\sqr78\,}\sqr{20.0}{18}\sqr{20.0}{18}} 
 
\def\spose#1{\hbox to 0pt{#1\hss}}\def\lta{\mathrel{\spose{\lower 3pt\hbox
{$\mathchar"218$}}\raise 2.0pt\hbox{$\mathchar"13C$}}}  \def\gta{\mathrel
{\spose{\lower 3pt\hbox{$\mathchar"218$}}\raise 2.0pt\hbox{$\mathchar"13E$}}}

\begin{document} 
 
\title{RELATIVISTIC SUPERFLUID MODELS FOR \\ ROTATING NEUTRON STARS}

\author{ \bf {Brandon Carter} 
\\ \\
Observatoire de Paris, 92195 Meudon, France.
\\ Email: {\it Brandon.Carter@obspm.fr}}

\date{Chapter contributed to \\
{\it PHYSICS OF NEUTRON STAR INTERIORS, \\
(Trento, June 2000)}, \\ed. D. Blasche, N.K. Glendenning, A. Sedrakian}
 
\maketitle  

\vskip 2 cm

\noindent{\bf Abstract}

This article starts by providing an introductory overview of the
theoretical mechanics of rotating neutron stars as developped to
account for the frequency variations, and particularly the
discontinuous glitches, observed in pulsars. The theory suggests,
and the observations seem to confirm, that an essential role is played 
by the interaction between the solid crust and inner layers whose 
superfluid nature allows them to rotate independently. However many
significant details remain to be clarified, even in much studied cases
such as the Crab and Vela.  The second part of this article is more 
technical, concentrating on just one of the many physical aspects that 
needs further development, namely the provision of a satisfactorily 
relativistic (local but not microscopic) treatment of the effects of 
the neutron superfluidity that is involved.

\vfill\eject

\section{Elementary global mechanics of rotating neutron stars}

\subsection{Introduction.}

Long before their observational detection as pulsars, theoreticiens
were well aware~\cite{WW} of the special physical interest of neutron 
stars -- whose existence was confidently predicted -- as well as of 
the (still entirely speculative) possibility of other more exotic 
(e.g. strange) stars of comparable compactness, meaning a radius only
a few times larger than the Schwarzschild limit value,
$R=2{\rm G} M/c^2$, for a mass comparable with that of our Sun.
Having presumably been formed by collapse of a stellar core that
marginally exceeds the Chandrasekhar limit for for a self gravitating
body with insufficient thermal pressure, a typical neutron star can be
expected to a have a mass rather close to this limit, which -- in terms 
of Newton's constant G, the speed of light $c$, the Dirac Planck 
constant $\hbar$, and the proton mass $m_{_{\rm p}}$ -- is given very 
roughly by the simple formula 
\begin{equation}\label{00M} M\approx \Big({\hbar c\over {\rm G}}
\Big)^{3/2} m_{_{\rm p}}^{\, -2} \, , \end{equation}
whose derivation is based just on the supposition that $m_{_{\rm p}}$
gives a rough estimate of the mass per cubic Fermi length, regardless
whether the degenerate relativistic fermions in question are
electrons (as in an ordinary white dwarf) neutrons, or even quarks.
 
Unlike what was possible when superfluidity of the neutron matter in 
such compact stars was originally predicted~\cite{Mig59} by Migdal,
present day article accelerators can explore the physics of individual 
particle at energies that are now approaching the order of a TeV. 
Nevertheless, although their levels -- from MeV to at most the order of 
GeV -- are only moderate by such modern standards, the thermal energies 
-- and particularly the Fermi energies -- characteristic of matter 
in neutron stars remain beyond the range accessible in the laboratory 
for bulk matter. 

For a mass near the value given by (\ref{00M}), the condition that 
the stellar radius be large compared with the Schwarzschild value, 
$R=2 {\rm G}M/c^2$, places an upper bound 
\begin{equation} \rho_\ast \ll \Big({c\over\hbar}\Big)^3 
m_{_{\rm p}}^{\,4}   \label{00d}\end{equation} 
on the mean stellar density $\rho_\ast$ -- and hence also on the central 
density (since unlike what is possible other kinds of stars, a neutron 
star cannot have a density profile that is sharply peaked at the center). 
While less compact neutron star configurations (with lower mass and 
larger radius) can exist in principle, it is hard to see how they could 
be created in nature, so a typical example can be expected to have a 
central density that is not so very far below what is permitted by 
this  Oppenheimer - Volkoff bound (\ref{00d}). Since this bound  is 
interpretable as the order of a proton mass per cubic proton Compton 
length, it is evidently quite a lot higher than the density of the order 
of a proton mass per cubic pion Compton length that characterises ordinary 
nuclear matter. In terms of the pion mass $m_{\pi}$ this ordinary 
nuclear density will be given in order of magnitude by
\begin{equation}\label{nuc} \rho_{_{\rm nuc}} \approx \Big({c\over\hbar}
\Big)^3 m_{\pi}^{\,3}m_{_{\rm p}} \, , \end{equation} 
which is a few times $10^{14}$ gm/cm$^3$. The prediction that typical 
neutron star core densities are thus well beyond what is easily accessible 
to experiment is one of the reasons why it is so interesting, not just 
for astronomy, but also for the basic physics~\cite{Glenden97,Weber99}
of bulk matter at the corresponding intermediate energy levels, to acquire 
and analyse as much observational information as possible about neutron 
stars (as well as ``strange'' or other comparably compact stars, which, if 
they exist, will also have core densities in the same range). 

\begin{figure} 
\centering
\epsfig{figure=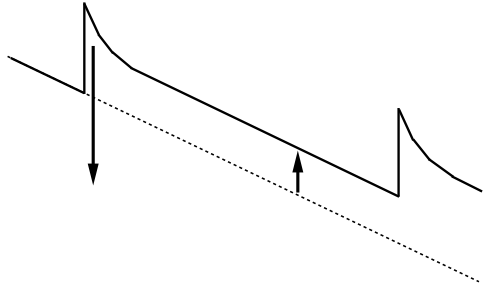, width=2.5 in} 
\caption{Qualitative sketch of a typical observational plot of pulsar 
angular velocity $\Omega$ against time $t$. The long down - pointing
arrow indicates the negative change $\Delta\Omega$ during a period of
steady slow down. The short up - pointing arrow indicates (on an 
exagerated scale) the positive jump $\delta\Omega$ during a glitch 
(consisting of a sharp discontinuity followed by a transient 
readjustment). 
\label{figC1}} 
\end{figure}

In addition to the limited amount of such information that is available 
from other mechanisms (such as binary orbital behaviour), we are fortunate 
to have at our disposal an enormous and steadily increasing body of 
relevant information provided ( see Figure~\ref{figC1}) by pulsar timing 
measurements: radio (and in some cases optical or other) observations 
provide continuous high (sometimes within $10^{-9}$) precision monitoring 
of pulsar frequencies, which are generally believed to correspond 
directly to the rotation frequency $\Omega$ of the underlying star, 
or more precisely to that of its rigidly rotating outer ``crust''.

The present article (like a briefer preceeding review~\cite{L00}) 
is intended as a self contained introduction to the theory of the 
phenomena most relevant to such observations. It is meant to be 
accessible to non-specialist readers, who are assumed just to have 
a grounding in general physics, at the level provided by Landau and 
Lifshitz~\cite{LL59}, in areas including relativistic gravitation theory, 
and non relativistic superfluidity and superconductivity theory.

As discussed in detail in accompanying articles in this volume, outside a 
still mysterious core (that may consist of quark matter) neutron stars 
are generally believed to consist essentially of a neutron fluid interior 
and a surrounding crust. The outer crust material is qualitatively similar 
to an ordinary metal, consisting of baryons concentrated (as a majority of 
neutrons with a minority of protons) in nuclei in a degenerate Fermi type 
sea of electrons at concentrations up to and beyond the white dwarf limit, 
where the electrons become relativistic, at a density given in terms of 
the electron mass $m_{\rm e}$ by
\begin{equation}\label{rel} \rho_{\rm rel} \approx \Big({c\over\hbar}\Big)^3 
m_{\rm e}^{\,3}m_{_{\rm p}}\, ,   \end{equation} 
whose value, of the order of $10^7$ gm/cm$^3$, corresponds to about one 
proton mass per cubic electron Compton length. 

The transition to the qualitatively different kind of material that makes the
behaviour of neutron stars so very different from that of ordinary degenerate 
electron supported white dwarf stars occurs at a critical ``neutron drip'' 
density $\rho_{\rm drip}$ that is reached when the Fermi energy of the 
degenerate electrons becomes comparable the binding energy $E_{\rm nuc}$ 
per baryon in a nucleus, whose value is of the order of the Fermi energy of 
the protons and neutrons when their mean separation is of the order of a pion 
Compton length, i.e.  $E_{\rm nuc}\approx (m_\pi c)^2/2 m_{_{\rm p}}$. 
Above this density,  
\begin{equation}\label{drip}  \rho_{\rm drip} \approx \Big({c\over\hbar}
\Big)^3  \Big({m_\pi^{\,3}\over 2 m_{_{\rm p}}}\Big)^2\, ,\end{equation}
which works out to be a few times $10^{11}$ gm/cm$^3$, the crust matter will 
still contain positively charged baryonic nuclei in a negatively charged Fermi 
sea of electrons, but there will now also be a third constituent consisting of 
freely moving neutrons outside the nuclei.

The use of the term ``crust'' to describe the layers both above and below the
critical value (\ref{drip}) is motivated by the consideration that the ionic
nuclei  will crystallise as a Coulomb lattice whose large scale behaviour 
will be that of an elastic solid as soon as the star has cooled sufficiently. 
Except for a very thin outer surface layer with density below the white dwarf
limit (\ref{rel}) that may remain liquid as a relatively shallow ``ocean'',
the rest of the crust is expected~\cite{BP71} to have solidified by the time 
the neutron star has dropped below the MeV level, which will be reached
within a few months of its formation. Due to the high conductivity of the 
degenerate electrons the outer magnetic field will be firmly anchored in
this crust, whose rotation rate is therefore what is  is measured directly
by pulsar frequency observations. 

For the purpose of explaining these observations (see Figure~\ref{figC1})
the most interesting feature of the crust is the presence of the
interprenetrating neutron fluid in the inner crust, at densities ranging
from the critical ``drip'' value (\ref{drip}) all the way up to the
nuclear value (\ref{nuc}) that is reached at the base of the crust beyond 
which the ions dissolve. In the relatively low temperatures (below an MeV) 
that are relevant it is generally believed~\cite{BPPR69,HGRR70} that 
the unconfined neutrons (like those within the nuclei)  will form Cooper 
type pairs that will form a bosonic condensate. The interpenetrating 
neutron constituent is thereby endowed with the property of superfluidity, 
which enables it to flow freely past the metallic lattice (and the
electron sea to which the lattice is electrically coupled) in the manner 
illustrated schematically in the following diagram (using hyphens to 
indicate the negatively charged electrons, crossed circles to indicate 
the positively charged nuclei, and double arrows to indicate relatively 
moving Cooper pairs of neutrons): 
\begin{eqnarray}\label{4}
- \ \ - \ \ - \ \ \Rightarrow \ \  -  \ \  -  \ \  -  \ \  \Rightarrow \  \
- \ \ - \ \ - \ \ \Rightarrow \ \  -  \ \  -  \ \  -  \ \  \Rightarrow \  \
- \ \ - \ \ -  \, \nonumber\\
- \ \ \Rightarrow \ \  -  \ \  -  \ \  -  \ \  \bigoplus\, \  - \ \ - \ \
- \ \ \Rightarrow \ \  -  \ \  -  \ \  -  \ \   \bigoplus\,   \  - \ \ - \ \
- \ \ \Rightarrow \ \  -  \nonumber\\
- \ \ - \ \ - \ \ \Rightarrow \ \  -  \ \  -  \ \  -  \ \  \Rightarrow \  \
- \ \ - \ \ - \ \ \Rightarrow \ \  -  \ \  -  \ \  -  \ \  \Rightarrow \  \
- \ \ - \ \ -  \, \nonumber\\
- \ \ \Rightarrow \ \  -  \ \  -  \ \  -  \ \  \Rightarrow\  \  - \ \ - \ \
- \ \ \Rightarrow \ \  -  \ \  -  \ \  -  \ \  \Rightarrow\  \  - \ \ - \ \
- \ \ \Rightarrow \ \  -  \nonumber\\
- \ \ - \ \ - \ \ \Rightarrow \ \  -  \ \  -  \ \  -  \ \  \Rightarrow \  \
- \ \ - \ \ - \ \ \Rightarrow \ \  -  \ \  -  \ \  -  \ \  \Rightarrow \  \
- \ \ - \ \ - \, \nonumber\\
- \ \ \bigoplus \, \  -  \ \  -  \ \  -  \ \  \Rightarrow\  \  - \ \ - \ \
- \ \ \bigoplus \, \  -  \ \  -  \ \  -  \ \  \Rightarrow\  \  - \ \ - \ \
- \ \ \bigoplus \, \  -  \nonumber\\
- \ \ - \ \ - \ \ \Rightarrow \ \  -  \ \  -  \ \  -  \ \  \Rightarrow \  \
- \ \ - \ \ - \ \ \Rightarrow \ \  -  \ \  -  \ \  -  \ \  \Rightarrow \  \
- \ \ - \ \ - \, \nonumber\\
- \ \ \Rightarrow \ \  -  \ \  -  \ \  -  \ \  \Rightarrow\  \  - \ \ - \ \
- \ \ \Rightarrow \ \  -  \ \  -  \ \  -  \ \  \Rightarrow\  \  - \ \ - \ \
- \ \ \Rightarrow \ \  -  \nonumber\\
- \ \ - \ \ - \ \ \Rightarrow \ \  -  \ \  -  \ \  -  \ \  \Rightarrow \  \
- \ \ - \ \ - \ \ \Rightarrow \ \  -  \ \  -  \ \  -  \ \  \Rightarrow \  \
- \ \ - \ \ -  \, \nonumber\\
- \ \ \Rightarrow \ \  -  \ \  -  \ \  -  \ \  \bigoplus\, \  - \ \ - \ \
- \ \ \Rightarrow \ \  -  \ \  -  \ \  -  \ \   \bigoplus\,   \  - \ \ - \ \
- \ \ \Rightarrow \ \  -  \nonumber\\
- \ \ - \ \ - \ \ \Rightarrow \ \  -  \ \  -  \ \  -  \ \  \Rightarrow \  \
- \ \ - \ \ - \ \ \Rightarrow \ \  -  \ \  -  \ \  -  \ \  \Rightarrow \  \
- \ \ - \ \ - \, \nonumber
\end{eqnarray}
The unconfined neutrons thus constitute a massive component that can
rotate independently of the crust, thereby -- as explained below -- 
providing the most promising kind of mechanism for explaining the 
observed pulsar frequency glitches (see Figure~\ref{figC1}).

At the base of the inner crust, at densities above the nuclear value given
by (\ref{nuc}), it is generally believed that the neutron fluid and ionic 
constituents merge to form a uniform fluid composed mainly of (superfluid) 
neutrons but with an independently moving (superconducting) protonic 
constituent, and with the further complication~\cite{SSS82} that 
instead of forming ordinary scalar Cooper type pairs the neutrons at 
this deeper level condense as pairs of spin 1. At even higher densities, 
beyond that of ordinary nuclear matter, various more or less exotic 
possibilities have been suggested, but no firm concensus has yet emerged. 
For example Glendenning has predicted~\cite{G92,G95} that there will be a 
hybrid zone in which negatively charged drops of quark matter will condense 
within the surrounding positively  charged baryonic liquid, and moreover 
that they will crystallise to form an ionic solid analogous to that of the 
crust. At even higher densities, as the maximum allowed by (\ref{00d}) 
is approached, one might expect that there would be an inner core where 
the drops merge to form another homogeneous superconducting superfluid 
zone, that (unlike the outer, baryonic core) would be constituted purely 
of quark matter, in which interesting new kinds of superfluidity and 
superconductivity~\cite{ARW98,BSS99,ABR99,SBSV00} could occur. Like the
somewhat better understood inner crust and outer core regions, these
very high density inner regions may also be relevant to the interpretation of
pulsar frequency observations such as are illustrated in Figure~\ref{figC1}.

The overall situation is not just that the global structure and behaviour
of a neutron star is rather complicated, but furthermore that (as {\it a 
fortiori} for strange stars if they exist) many important aspects
remain so unclear that -- except in the crust region (or for very low 
mass neutron stars) for which a reasonable degree of concensus already 
prevails -- it is hardly worthwhile yet even to start the detailed 
numerical calculations that will be needed later on. Before a convincingly 
realistic neutron star model can be developed even as a rough approximation 
many underlying physical issues will need to be dealt with, of which the most 
basic are concerned with the qualitative nature of matter at the supernuclear 
densities attained in the cores of all neutron stars above or near the precise 
Chandrasekhar limit value, $M\simeq \sqrt 2\, M_{_\odot}$ (where $M_{_\odot}$ 
is the solar mass) though not for very low mass neutron star configurations 
(which are at least of academic interest, even though it is hard to see how 
they could be created in nature. 

From the point of view of the interpretation of observational data, many 
less fundamental but technically non-trivial issues need to be be clarified.
Among the other accessory issues  (concerning matter at less extreme 
densities) that also need to be dealt with, the one with which the 
present article will be primarily concerned is that of the consequences
of the predicted superfluidity. 
The final sections will concentrate on the results of recent progress on 
the development of an appropriately relativistic treatment as an 
improvement (in view of immediate coherence, as well
as the long term objecive of precision) over the non-relativistic
treatment that has until now been mainly used, not just for 
superfluidity but also for many other relevant phenomena.
However before going into the technical aspects of the relativistic 
treatment, the first part of this article will provide a brief
survey of the reasons why the phenomenon of superfluidity is
particularly important for relating theoretical understanding
of the inner structure of the neutron star to the available
observational data, of which the most richly informative part
(see Figure~\ref{figC1}) comes from pulsar timing.

\subsection{\bf Minimal two component rotating star models}

As emphasised above, for many of the most important questions about
the global structure of neutron stars, no quantitative agreement is 
available or to be expected for a long time yet. There are however 
several essential qualitative features on which practically all neutron 
star theorists do seem to agree already. In relation to the pulsar timing 
observations, the most important of these agreed features~\cite{BP71}
is the presence of a rigidly corotating structure that may or may not 
include part of the core but certainly includes a solid outer crust to 
which is anchored the magnetic field configuration that gives rise to 
the observed radio and other emission. The next most important feature, 
common to all viable theoretical scenarios  albeit of a more subtle 
nature, is the presence of some (maybe many) effectively superfluid 
(or superconducting) zones that can rotate independently~\cite{BPPR69} of 
the rigidly rotating crust structure whose angular velocity, 
$\Omega_{\rm c}$ say, is presumed to be the same as the $\Omega$ that is 
directly observed. 

The mimimal agreement about these two essential features is what 
underlies the longstanding, widespread, and enduring popularity of 
a corresponding kind minimally complicated rotating neutron star model, 
involving just two independently rotating parts: a ``corotating crust'' 
part with (directly observed) angular velocity $\Omega_{\rm c}$  and a 
``superfluid neutron'' part with a possibly different angular velocity 
$\Omega_{\rm n}$ (representing the average of what in a more detailed 
treatment would be a spacially variable angular velocity distribution).

The basic postulate of such a minimal model is that the total angular
momentum $J$ of the star is the sum of decoupled parts,
\begin{equation}\label{000}J=J_{\rm c}+J_{\rm n}\, , \end{equation}
 with 
\begin{equation}\label{001} J_{\rm c}= I_{\rm c}\Omega_{\rm c}\, ,
\hskip 1 cm J_{\rm n}= I_{\rm n}\Omega_{\rm c} \, ,\end{equation}
where $I_{\rm c}$ and $I_{\rm n}$ are separate moments of inertia that are 
supposed to remain constant during a process of continued variation 
governed by an external torque $\Gamma_{\rm ex}$. (A more sophisticated 
variant of this model would allow for cross coupling, whereby $J_{\rm c}$ 
is affected by $\Omega_{\rm n}$ and vice versa: a small cross coupling of 
this kind would inevitably be present~\cite{C74} in an exactly 
relativistic description, and a possibly more important cross coupling 
effect is to be expected from the effect -- to be discussed below 
-- of superfluid momentum ``entrainment'', whose likely relevance in 
neutron star matter was originally pointed out in the context of proton 
superconductvity~\cite{VS81,ALS84}.)

The external couple $\Gamma_{\rm ex}$ represents the effect of the 
magnetic field anchored in the rotating crust, which, in view of the high 
conductivity of the crust is generally assumed to remain constant over 
timescales long compared with those (at most a few years, since the oldest 
pulsar observations go back only to 1968) of the observed fluctations. 
The effect of this steady couple is of course to cause a total angular 
momentum loss rate given by
\begin{equation}\label{002} \dot J=\Gamma_{\rm ex} < 0\, . \end{equation}
If the angular velocities were locked together,
$\Omega_{\rm n}=\Omega_{\rm c}=\Omega$,  this would  give
$\dot\Omega=\Gamma_{\rm ex}/I$ with 
\begin{equation}\label{003}I=I_{\rm c}+I_{\rm n}\, ,\end{equation} 
and this relation will in any case be true for the long term average, 
$\langle\Omega\rangle$ i.e. since (for reasons to be discussed below) the 
separate angular velocities can never get too far appart, the long term 
slowdown of the observed pulsar frequency allows the torque involved to 
be estimated as 
\begin{equation}\label{004} \Gamma_{\rm ex}
=I\,\langle \dot \Omega\rangle\, . \end{equation}
in which, for neutron stars with mass $M\simeq3 M_{_\odot}/2$ (which, 
consistently with the theoretical prediction (\ref{00M}), is what has 
been found~\cite{TC99} for the few cases in which the mass is 
reliably measurable), the total moment of inertia (unlike the 
distinct parts $I_{\rm n}$ and $I_{\rm c}$ for which different 
theoreticians have rather diverse ideas in various cases) can be 
evaluated in a generally agreed manner, which leads~\cite{CGPB99} to 
estimates of about $10^2 M_{_\odot}$ Km$^2$ within a factor of order 
unity (whereas the uncertainty range would be much larger for a neutron 
star nearer the upper mass limit).

The idea of the two component model is that as well as supporting the
effect of the external torque $\Gamma_{\rm ex}$, the crust component
exerts an internal torque $\Gamma_{\rm in}$ on the ``superfluid neutron''
component, which therefore obeys an evolution equation of the form
\begin{equation}\label{005} \dot J_{\rm n}=\Gamma_{\rm in}\, ,\end{equation}
while, in order to be consistent with (\ref{003}) the crust component must 
obey
\begin{equation}\label{006} \dot J_{\rm c}=
\Gamma_{\rm ex}-\Gamma_{\rm in} \, ,\end{equation}
in which, unlike the external torque $\Gamma_{\rm ex}$, the internal torque 
is not constant but proportional to the angular velocity difference
\begin{equation}\label{007} \Gamma_{\rm in}= -{I_{\rm n}I_{\rm c}\over I}\, 
{\omega\over\tau} \, ,\hskip 1 cm \omega=\Omega_{\rm n}
-\Omega_{\rm c}\, , \end{equation} 
where $\tau$ is a damping timescale whose estimation will be discussed
below. The chosen normalisation of this timescale is such that, according 
to the preceeding equations, the angular velocity difference will satisfy a 
differential equation of the simple form
\begin{equation}\label{008}
\dot\omega +{\omega \over\tau}=- {I\over I_{\rm c}}
\langle \dot \Omega\rangle\, , \end{equation}
in which the right hand side is a constant that fixes the saturation  limit
\begin{equation}\label{009} \omega\rightarrow \omega_{\rm s}\, , 
\hskip 1 cm \omega_{\rm s}= - {I\over I_{\rm c}}\,
\langle \dot \Omega\rangle\, \tau > 0\, ,\end{equation}
to which the difference $\omega$ will tend in the long run, unless 
this continuous evolution process is interrupted by a ``glitch''.
So long as no such interruption  has occurred, the angular velocity 
difference will be given as a function of the time $t$ by an 
expression of the form
\begin{equation}\label{017}  \omega=\omega_{\rm s} + 
(\omega_0 -\omega_{\rm s}) {\rm e}^{- t/\tau}\, ,\end{equation}
where $\omega_0$ is a constant of integration interpretable as the
value of $\omega$ when the clock time $t$ was set to zero.

\subsection{ The problem of accounting for glitches}

Some of the strongest observational evidence in favour of the 
theoretical picture epitomised by the kind of highly simplified 2 
component neutron star model described in the previous section 
is provided by a phenomenon in which the continuous evolution 
described by this model is subject to a temporary model break down. 
The phenomenon in question is what is known as a glitch (see 
Figure~\ref{figC1}) during which (with a rise time too short to be 
measured, at most a few hours and probably much less) the observed 
angular frequency $\Omega$ of an isolated pulsar undergoes a 
positive discontinuity, $\delta\Omega>0$, that partially cancels 
the loss $\Delta\Omega$ during the preceeding period of continuous 
slowdown. 

Since there is no imaginable way the external torque could suddenly 
become very large (nor any observational evidence that the 
associated pulsar emission process changes significantly at all 
during a glitch) there can be no corresponding discontinuity in the 
angular momentum. This means that, if we want to use a model only
involving a single component, it will be necessary to take account
of variation of the moment of inertia, whose total $I$ will 
evidently undergo a negative variation $\delta I$ given, for
a glitch of amplitude $\delta\Omega$, by
\begin{equation}\label{010} \delta J=0 \ \Rightarrow \ \delta I=-I\, 
{\delta \Omega\over\Omega} \, .\end{equation}

The earliest theory designed to account for this phenomenon, as first 
observed in the Crab and Vela pulsars, was based on the first of what 
were presented in the preceeding section as generally agreed features 
of neutron stars, namely the presence of a solid crust structure, but 
on the basis of a single component model taking no account of the 
second generally agreed feature (namely the possibility of 
independently rotating parts). The idea~\cite{R69,BPPR69,S70} was that 
the rigidity of the solid crust would tend to prevent the decrease
in moment of inertia that would necessarily accompany the loss of 
angular momentum in a purely fluid star. In a simple rotating 
fluid star model, the oblateness due to centrifugal force would 
give rise to a variable moment of inertia that would be expressible 
for low values of the angular velocity by an expression of the 
quadratic form
\begin{equation}\label{011} I\simeq I_0\Big(
1+{\Omega^2\over\Omega_\ast^{\,2}}\Big)\, ,\end{equation}
where $I_0$ is the value of the moment of inertia in the
non-rotating spherical limit and $\Omega_\ast$ is a constant
specifying the relatively high value of the angular velocity (which 
will be given in terms of the mean density subject to (\ref{00d}) 
by the rough order of magnitude estimate   
$\Omega_\ast^{\,2}\approx$ G$\,\rho_\ast$) that would be necessary 
for deviations from spherical symmetry to be of order unity. Whatever 
may have happened immediately after the birth of the neutron star, no 
such rapid rotation still occurs in any of the (at least centuries 
old) pulsars that are actually observed today, for which the condition 
$\Omega^2 \ll\Omega_\ast^{\,2}$ is always satisfied.

For a simple perfect fluid star model, the effect of the external
torque (\ref{002}) during an extended time interval $\Delta t$ would 
be to cause an angular momentum loss,
$\Delta J\simeq \Gamma_{\rm ex} \Delta t <0$
that would be accompanied by a corresponding angular velocity 
variation $\Delta\Omega<0$, which according to (\ref{011}) 
would entail a decrease in  moment of inertia given by
\begin{equation}\label{012} \Delta I \simeq 
2I \,{\Omega^2\over\Omega_{\ast}^{\, 2}}
{\Delta\Omega\over\Omega}<0\, . \end{equation}
Due to the solidity of the crust, which tends to preserve the more 
highly elliptic initial configuration, the actual change in the moment 
of inertia will fall short of what is predicted by this formula, but at 
some stage the strain will build up to the point at which the solid 
structure will break down. The implication is that there will then be 
a ``crustquake'', in which the solid structure suddenly changes towards 
what the perfect fluid structure would have been, thereby changing the 
moment of inertia by an amount that will be at most of the order of the 
upper limit given by  
\begin{equation}\label{013}
 \delta I\lta \Delta I,\end{equation}
where $\Delta I$ is what is given by (\ref{012}), and that will 
be considerably less than this if the crust rigidity is low.
According to (\ref{010}) the corresponding positive angular velocity
discontinuity $\delta\Omega$ associated with the continuous
negative angular velocity change $\Delta\Omega$ since the
preceeding glitch will be subject to the limit
\begin{equation}\label{015} \delta\Omega \lta -2 \,{\Omega^2\over
\Omega_{\ast}^{\, 2}}\, \Delta\Omega\, . \end{equation}

The preceeding formula provides an order of magnitude limit that 
must be satisfied by a rather large margin if the rigidity is 
low but that is entirely consistent with what is observed in the case 
of the Crab pulsar, for which typical glitches are characterised by 
$\delta \Omega\approx 10^{-8}\Omega$. However almost immediately 
after it was first proposed, it began to be recognised~\cite{BP71}  
that this rather obvious mechanism would not be sufficient to account 
for the much larger glitches that are frequently observed in cases 
such as that of the Vela pulsar, for which typical glitch
amplitudes are characterised by 
$\delta\Omega\approx 10^{-6}\Omega$ $ \approx -10^{-2} \Delta\Omega$.
 
Since it was first suggested by pioneers such as Anderson and 
Itoh~\cite{AI75}, the  generally accepted way of getting round this 
limitation -- namely that the likely changes of the moments of inertia 
will be far too small to account for frequent giant (Vela type) 
glitches -- is to drop the single component description in favor of 
the two component description in which glitches can be accounted for 
even if (as assumed in its simplest version) the relevant moments of 
inertia undergo no significant change at all. The essential point is 
that the consideration that the very short glitch duration excludes 
any significant jump in the total angular momentum does not rule out 
the possibility of impulsive transfer of angular momentum between the 
two components provided they balance out:
\begin{equation}\label{018} \delta J=0\ \ \ \Rightarrow \ \ \ 
\delta J_{\rm c}=-\delta J_{\rm n}\, ,\end{equation}
so that
\begin{equation}\label{016} \delta\Omega_{\rm n}= 
-{I_{\rm c}\over I_{\rm n}}\,\delta\Omega_{\rm c}\, .\end{equation}
The idea is that between the glitches the weak coupling mechanism
described by (\ref{008}) allows the slowdown of the ``neutron superfluid'' 
angular velocity $\Omega_{\rm n}$ to lag behind that of the crust 
component which is what is presumed to be actually observed, 
$\Omega=\Omega_{\rm c}$, so that during the preceeding period 
$\Delta t$ the angular velocity difference $\omega$ will be positive.
It is generally supposed that (for diverse reasons to be discussed 
below) this angular velocity difference gives rise to stresses that 
are partially relaxed in the glitch process, whose onset occurs
when the difference $\omega$ reaches a critical glitch inducing
value $\omega_{\rm g}$ say that, in order to be attainable must be 
less than the limit $\omega_{\rm s}$ given by (\ref{009}) -- a condition 
that would fail if the relaxation timescale $\tau$ were too short.
 
Leaving aside cases for which $\omega_{\rm g}>\omega_{\rm s}$
(whose evolution will be of the glitch free kind recently 
investigated~\cite{CS98} by Sedakian and Cordes) as well as 
the marginal case in which $\omega_{\rm g}\approx 
\omega_{\rm s}$, i.e. subject to the proviso that there is a safe 
margin $\omega_{\rm g}\ll \omega_{\rm s}$, the evolution equation 
(\ref{017}) will be replacable by the linear relation
\begin{equation}\label{020} {\omega\over\omega_{\rm s}}=
{\omega_0\over\omega_{\rm s}}+{t\over\tau} \, ,\end{equation}
in which each of the terms is small compared with unity. 
Successive glitches bring about negative adjustments $\delta\omega$
that are needed to cancel out the cumulative effect of the
positive variations $\Delta$ that develop during the duration of the 
interglitch periods governed by (\ref{020}), so that on average
they cannot deviate too much from the order of magnitude estimate 
given simply by
\begin{equation}\label{021} \delta\omega\approx 
-\Delta\omega\, , \end{equation}
in which, by (\ref{009}) and (\ref{020}), the deviation built up 
during an interglitch interval of duration $\Delta t$ will be given 
simply by
\begin{equation}\label{022} \Delta\omega= -{I\over I_{\rm c}}
\langle\dot\Omega\rangle \Delta t \, .\end{equation}
Using (\ref{016}) to eliminate the unobservable jump 
$\delta\Omega_{\rm n}$ from the difference $\delta\omega$ 
$=\delta\Omega_{\rm n}-\delta\Omega_{\rm c}$, the magnitude of the 
observable jump $\delta\Omega_{\rm c}$ can be estimated by (\ref{021}) 
as
\begin{equation}\label{023}\delta\Omega_{\rm c}\approx 
{I_{\rm n}\over I} \,\Delta\omega\, .\end{equation}
Since the observable interglitch frequency variation will be given 
roughly by $\Delta\Omega_{\rm c}\simeq
\langle\dot\Omega\rangle\Delta t$ one sees from (\ref{022}) and 
(\ref{023}) that it provides a corresponding estimate
\begin{equation}\label{024}\delta\Omega_{\rm c}\approx 
-{I_{\rm n}\over I_{\rm c}} \Delta\Omega_{\rm c}\, ,\end{equation}
for the observable frequency jump during a glitch. The presumption 
that $\Omega_{\rm c}$ is identifiable with the $\Omega$ that is 
observed allows us to compare this with the previous upper limit
(\ref{015}) that was obtained for the single component model 
with variable moment of inertia. It can be seen that the difference 
is simply that the small factor $(\Omega/\Omega_\ast)^2$ in
the upper limit for the single component model is replaced, 
in the two component fixed moment of inertia model, by the ratio 
$I_{\rm n}/I_{\rm c}$ whose value is highly uncertain (in view of our 
lack of firm knowledge about what goes on in the core of the neutron
star) but can plausibly be supposed to be of the order of unity, which 
is what is needed to account for the frequent very large glitches
(with $\delta\Omega\approx -10^{-2}\Delta\Omega$) that 
are observed in examples such as Vela.

\subsection{The question of pinning and the damping timescale}

The foregoing estimate (\ref{024}) is not sensitive to the particular
value of the damping timescale $\tau$ except that it is assumed to be 
large compared with the interglitch period $\Delta t$ which is 
usually several months or more. This requirement might at first sight 
appear to be incompatible with observations of post glitch relaxation, 
in which shorter timescales of only a few weeks have been shown to be 
involved. Such discrepancies are however to be expected on the basis 
of our general qualitative understanding~\cite{AAPS81} of what is 
involved. The strong density gradients in the star imply the existence 
of many different zones in which differential rotation with a wide 
range of damping timescales can occur. Our simplified two component 
description of glitches depends on taking the part with moment of 
inertia $I_{\rm n}$ to correspond to a substructure for which the 
relevant damping timescale timescale $\tau$ is very large. A formally 
similar two constituent model might also be used for describing post 
glitch relaxation with much shorter timescales, but for such an 
application the substructure with moment of inertia $I_{\rm n}$ would 
have to be reinterpreted as corresponding to some other part of the 
star. Of course if we wanted to describe both the glitches and the
postglitch relaxation, in a single coherent framework,  we would
need to amalgamate the separate two component models so as to obtain 
a more elaborate model (such as has recently been used~\cite{SWC99}  
for the analysis of precession) with three or more independent 
components (and with not just a single damping timescale but an 
antisymmetric matrix of mutual damping coefficients). Although the 
construction of such composite models is straightforward in principle, 
most authors have so far (quite reasonably) preferred to concentrate on 
particular aspects for which a less intricate description is adequate.

Even for applications, such as the glitch model of the previous section,
for which a two component description is adequate as a lowest
order approximation, the estimation of the relevant damping timescale
remains a subject of great uncertainty. The situation has however
been clearer since the general question of quasi stationary equilibrium
in a rotating superfluid was systematically addressed in the context
of neutron stars by Alpar and Sauls~\cite{AS88}, Bildsten and 
Epstein~\cite{BE89} and the Sedrakians~\cite{SS92}, who drew attention
to the consideration that long damping timescales can arise not just from
weak but also from strong coupling. These authors considered the basic 
general problem of a two constituent superfluid model of the simplest
kind in a local configuration of differential rotation about a fixed axis 
characterised by a unit 3 vector, $\vec \nu$ say in the neighbourhood of 
a position determined (in a Newtonian flat space description) with 
respect to a central rotation axis by an orthogonal radius vector 
$\vec r$. One of the constituents is the ``normal'' (and therefore in a
n equilibrium state) rigidly rotating) crust constituent characterised 
by a (uniform) angular velocity $\Omega_{\rm c}$ and a corresponding 
velocity vector given as the cross product $\vec v_{\rm c} $ 
$=\Omega_{\rm c}\, \vec\nu \times \vec r$. The large scale averaged 
velocity of the superfluid constituent -- which for our purpose is 
to be thought of as constituted of neutrons -- is characterised in 
terms of a perhaps radially variable angular velocity $\Omega_{\rm n}$ 
by a similar formula $\vec v_{\rm n} =\Omega_{\rm n}\,\vec\nu 
\times \vec r$. However in the latter case it is to be born in mind 
that that on a microscopic scale the superfluid fluid is irrotational 
except on quantised vortex lines round which the integral of the 
relevant superfluid particle momentum $m v_{\rm n}$ is given by the 
Planck constant, i.e. $2\pi\hbar$ in Dirac's notation, so that the 
corresponding velocity circulation is $\kappa=2\pi\hbar/m$  -- while, 
in view of the bosonic pairing, the relevant mass scale in the 
application with which we are concerned is twice that of the 
neutron, i.e.  $m=2 m_{\rm n}$

Since the averaged velocity circulation per unit area orthogonal to
the rotation direction $\vec\nu$ will simply be $2\Omega_{\rm n}$,
it follows that the corresponding surface number density, $\sigma$ say,
of quantised vortices will be given by 
\begin{equation}\label{029}
\sigma={\Omega m\over\pi\hbar}\, .\end{equation}
Although the superfluid motion is non dissipative, and so has no direct
interaction with the ``normal'' crust material, the vortex cores
(which are defects where the superflidity breaks down) will in general
be subject to a drag force $\vec F_{\rm d}$ say per unit length,
exerted by the bacground in the direction of relative motion.
Using the notation $\vec v_{\rm v}$ for the velocity of motion
of the vortex lines orthogonally to the rotationaxis $\vec\nu$
the dissipative drag force exerted by the background will be given
by a formula of the form
\begin{equation}\label{030} \vec F_{\rm d}=
\eta_{\rm r} \big(\vec v_{\rm c}-\vec v_{\rm v}\big)\, ,\end{equation}
in which $\eta_{\rm r}$ is a positive resistive drag coefficient, whose 
quantitative evaluation is a subject of much uncertainty -- not just in 
the core, but even in the crust, where it is very sensitive to 
temperature~\cite{AAPS84} and other quantities such as superfluid 
pairing correlation lengths that are rather difficult to 
estimate~\cite{J90}. 

This drag will not be the only force acting on the vortex 
line, which  will also be subject to the Magnus effect. 
According to the well known formula~\cite{LL59} of Joukowski
(or, in an alternative transliteration, Zhukovskii) this gives 
rise to a lift force $\vec F_{\rm l}$ per unit length that is  
proportional to the product of the particle number density
and the corresponding momentum circulation: 
\begin{equation}\label{031} \vec F_{\rm l}= 2\pi \hbar n \big(\vec v_{\rm n} 
- \vec v_{\rm v}\big) \times \vec\nu\, ,\end{equation}
where $n$ is the relevant particle number density. In the neutron 
superfluid application with which we are concerned the effect of the
bosonic pairing must be taken into account, which means that the 
relevant number density is only half that of the neutrons themselves,
i.e. $n=n_{\rm n}/2$.
       
On the microscopically very long timescales characterising the 
relevant applications the vortex lines can be treated as effectively 
massless, which means that the evolution of the system will be 
determined simply by the condition that the total force on a vortex
line must cancel out,
\begin{equation}\label{032} \vec F_{\rm d}+ \vec F_{\rm l}
=0\, .\end{equation}
To solve this, it is convenient to decompose the velocity $\vec v_{\rm v}$
of the vortex lines into a (small) radially outward directed part, with 
magnitude $\dot r$, and a (larger, but for our purpose less important)
remainder, directed parallel to the fluid flow vectors,
in the form
\begin{equation}\label{033} \vec v_{\rm v}={\dot r\over r}\,\vec r
+\Omega_{\rm v}\, \vec\nu \times \vec r\, ,\end{equation}
where $\Omega_{\rm v}$ is interpretable as the angular velocity of the
vortex lattice. It is also convenient to introduce a dimensionless
resistivity coefficient defined by
\begin{equation}\label{034} c_{\rm r}=
{\eta_{\rm r}\over 2\pi\hbar n}\, ,\end{equation}
which is what in the jargon of aero engineering would be called the
drag to lift ratio (what, in that context, one seeks to minimise by 
cunning aerofoil design). The solution of (\ref{033}) is thereby 
expressible as the condition that the vortex line angular velocity is 
intermediate between those of the crust and supefluid constituents, with 
value given by 
\begin{equation}\label{036} \Omega_{\rm v}={\Omega_{\rm n}+c_{\rm r}^{\,2}
\,\Omega_{\rm c}\over 1 + c_{\rm r}^{\,2}}\, ,\end{equation}
while the radially outward ``creep'' component of the velocity of the 
vortex lines will be given by
\begin{equation}\label{037} \dot r= {r c_{\rm r}\over 1 + 
c_{\rm r}^{\,2}}\,\big(\Omega_{\rm n}-\Omega_{\rm c}\big)
\, .\end{equation}

This last equation is particularly important because it determines 
the rate of change of the vortex line surface density $\sigma$: 
as the comoving radius of the vortex distribution increases, 
the surface density will evidently undergo a corresponding decrease 
given by the relation $\dot\sigma/\sigma=-2\dot r/r$. Since by 
(\ref{029}) this surface density is proportional to the superfluid 
angular velocity, we deduce that the rate of variation of the latter 
will be given by
\begin{equation}\label{038} {\dot\Omega_{\rm n}\over\Omega_{\rm n}}
= -{2 c_{\rm r}\,\omega \over 1 + c_{\rm r}^{\,2}}\, , \end{equation}
where $\omega$ is the angular velocity difference as introduced in the 
relation (\ref{007}), from which, by comparison with (\ref{038}), the 
corresponding value of the damping timescale $\tau$ in which we are 
interested in, can be read out as
\begin{equation} \tau\label{039}={I_{\rm c}\over 2 I\Omega_{\rm n}}
\Big(c_{\rm r}+{1\over c_{\rm r}}\Big) \, .\end{equation}

As well as showing that the timescale is subject to a lower limit
(that might have been guessed on dimensional grounds) given 
by $\tau\geq {I_{\rm c}/ I\Omega_{\rm n}}$ and attained for 
$c_{\rm r}=1$, a noteworthy feature of this 
result~\cite{AS88,BE89,SS92}, is the dual symmetry between the 
roles of the drag to lift ratio $c_{\rm r}$ and of its inverse, the
lift to drag ratio $c_{\rm r}^{-1}$. The decay timescale $\tau$ 
becomes infinitely large not just in the drag free limit for which 
$\eta8{\rm r}$ and hence $c_{\rm r}$ become arbitrarily small -- so that 
by (\ref{036}) the vortices are dragged allong with the superfluid -- 
but also in the opposite ``pinned'' limit of very large $c_{\rm}$, 
for which the force on the vortices is strong enough to lock them 
to the crust material.

The estimation of the actual values of the $c_{\rm r}$, in the various 
zones of interest, has been the subject of much work, but the subject is 
difficult and many of the results are still inconclusive or 
controversial. Following the recognition~\cite{AS88} that magnetic 
coupling forces between the crust and the superconducting proton neutron 
superfluid zone are more important than had been previously supposed, 
and strong enough to lock this core region to the crust on timescales
short compared with those relevant to glitch observations, it was 
suggested~\cite{SS95} that even the new coupling force values were 
underestimated, so much so that $c_{\rm r}$ would become large compared 
with unity, with the implication that $\tau$ could be very long after 
all, just as had been supposed in the early years when $c_{\rm r}$ had
been supposed to be small. Even in the qualitatively more familiar 
crust regime the situation is still unclear, partly because of effects 
of temperature dependence: work of Jones~\cite{J90,J91} suggests that 
pinning may be much less effective than had been previously supposed 
so that instead of being high $c_{\rm r}$ would be very low there.

\subsection{The long term crustal drift phenomenon}

The question of the effectiveness of vortex pinning to the crust
comonent leads on to the related issue of what is actually 
responsible for the stress whose release, when a critical value is 
exceeded, is supposed to provide the glitch mechanism in the two 
constituent scenario described in the preceeding section. All the early 
versions of such a two constituent mechanism assumed that the relevant
stresses would be due to vortex pinning. Like their more recent 
variants, the various early versions were classifiable in two 
distinct categories. In the first category ~\cite{AI75,LE91} it was supposed
that the discontinuous breakdown would occur when some maximum
static pinning force value was exceeded: the sudden (rather than 
``creeping'') nature of the breakdown was accounted for, in a recent 
version~\cite{LE96}, as being due to a thermal instability resulting 
from the temperature sensitivity of $c_{\rm r}$, while another new 
suggestion~\cite{SC99} is that the relevant slippage occurs at the locus
where the ions dissolve at the base of the crust. In the second
category~\cite{AI75,R76,R91} it was suggested that such a maximum
pinning force value might never be reached because the elastic solid 
structure would breakdown first in a crustquake (of the kind required 
in the single constituent moment of inertia changing mechanism that 
may account for cases such as that of the Crab). 

In all these various versions, the necessary transfer of angular momentum 
from the relevant independently rotating layers with moment of inertia 
$I_{\rm n}$ to the crust component with moment of inertia $I_{\rm c}$ 
is mainly attributable to the torque exerted by the pinning forces. 
However it has recently been pointed out~\cite{CLS00} that there is an 
alternative possibility (effectively a new variant within the second 
category) whereby the necessary angular momentum transfer 
may be acheived convectively -- by a transfer of matter (removal from the 
crust of matter with low angular momentum, and its replacement by 
matter with higher angular angular momentum) which can occur even if 
torque forces are entirely absent, i.e. in the small $c_{\rm r}$ limit.
In this new kind of scenario, the stress ultimately responsible (when a 
critical level has been exceeded) for the discontinuous transfer is 
attributable to a centrifugal buoyancy deficit in the relatively slowed 
down crust component.

\begin{figure} 
\centering
\epsfig{figure=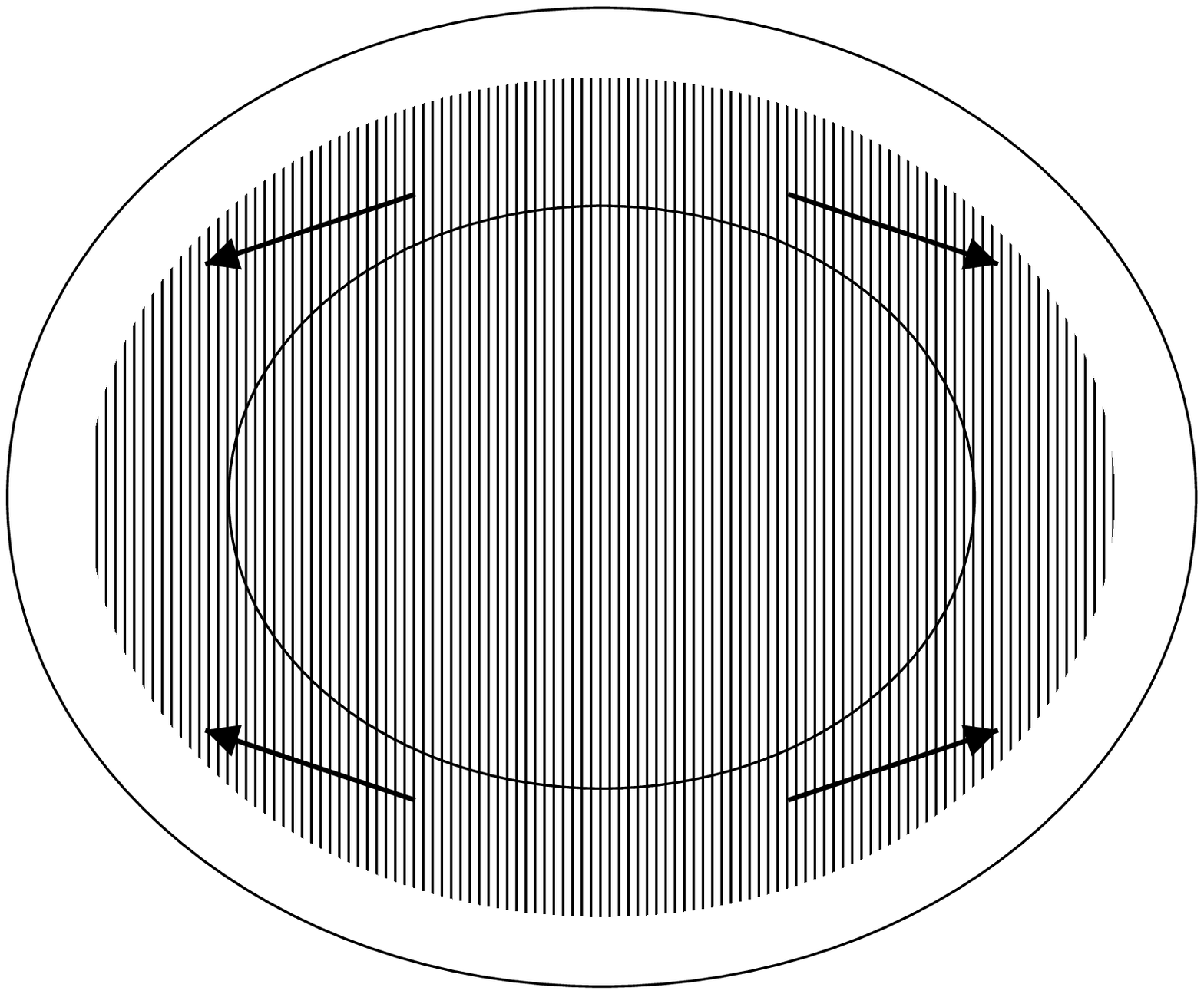, width=2.5 in} 
\caption{Qualitative sketch indicating direction of force expected 
(c.f Ruderman 1991) to act on (magnetically slowed down) crust  
due to vortex pinning mechanism, if it is  effective,  when the 
(interpenetrating) neutron superfluid retains a higher rotation rate. 
(Vertical shading indicates the alignement of the vortices in the region 
occupied by neutron superfluid, which is not confined  to the core but 
interpenetrates the greater part of the solid cust as well.) 
\label{figC2}} 
\end{figure}

In the case of the earlier pinning driven mechanism, it was pointed
out by Ruderman~\cite{R91} that if the glitches were due to breakdown
of the solid structure (rather than discontinuous vortex slippage) then 
the long term effect of many glitches would be analogous to that of 
terrestrial contentinental drift. It would give rise to a pattern of
convective circulation~\cite{P99} involving ``transfusion'' of matter 
from the crust constituent to the underlying neutron superfluid 
constituent in a ``subduction'' region near the equator, at colatitude 
$\theta=\pi/2$, and the other way round near the poles at colatitude 
$\theta=0$. The corresponding long term average rate 
$\langle\dot\theta\rangle$ of angular drift of a crust plate at the 
surface, which for the spin down of an isolated pulsar would be directed 
away from the pole towards the equator (see Figure~\ref{figC2}), was 
estimated by Ruderman on the assumption that it would correspond to an 
outward velocity of the same order of magnitude as the mean cylindrical 
expansion rate,
$ \langle\dot r\rangle\approx -\langle\dot\sigma\rangle/ 2\sigma$
of the vortex distribution, whose surface number density $\sigma$ is
given in terms of the angular velocity by the proportionality
relation (\ref{029}). This reasonning~\cite{R91} provided
a formula of the form
\begin{equation}\label{040} \langle\dot\theta\rangle\approx -{\langle\dot
\Omega\rangle\over 2\Omega}\, ,\end{equation}
which implies that the timescale for complete turnover of the crust 
material is of the same order as the spin down lifetime of the pulsar, 
during which, as Ruderman pointed out, the magnetic dipole would be able 
to be dragged most of the way from the rotation axis to the equator. 
This would result in a net increase (ocurring discontinuously at the 
glitches) of the pulsar radiation rate and thus of the magnitude of the 
spin down rate $\dot\Omega$. Another reason for an increase of the spin 
down rate would be the decrease in oblateness according to (\ref{011}), 
but this would evidently be much less important.  

\begin{figure}
 \centering
\epsfig{figure=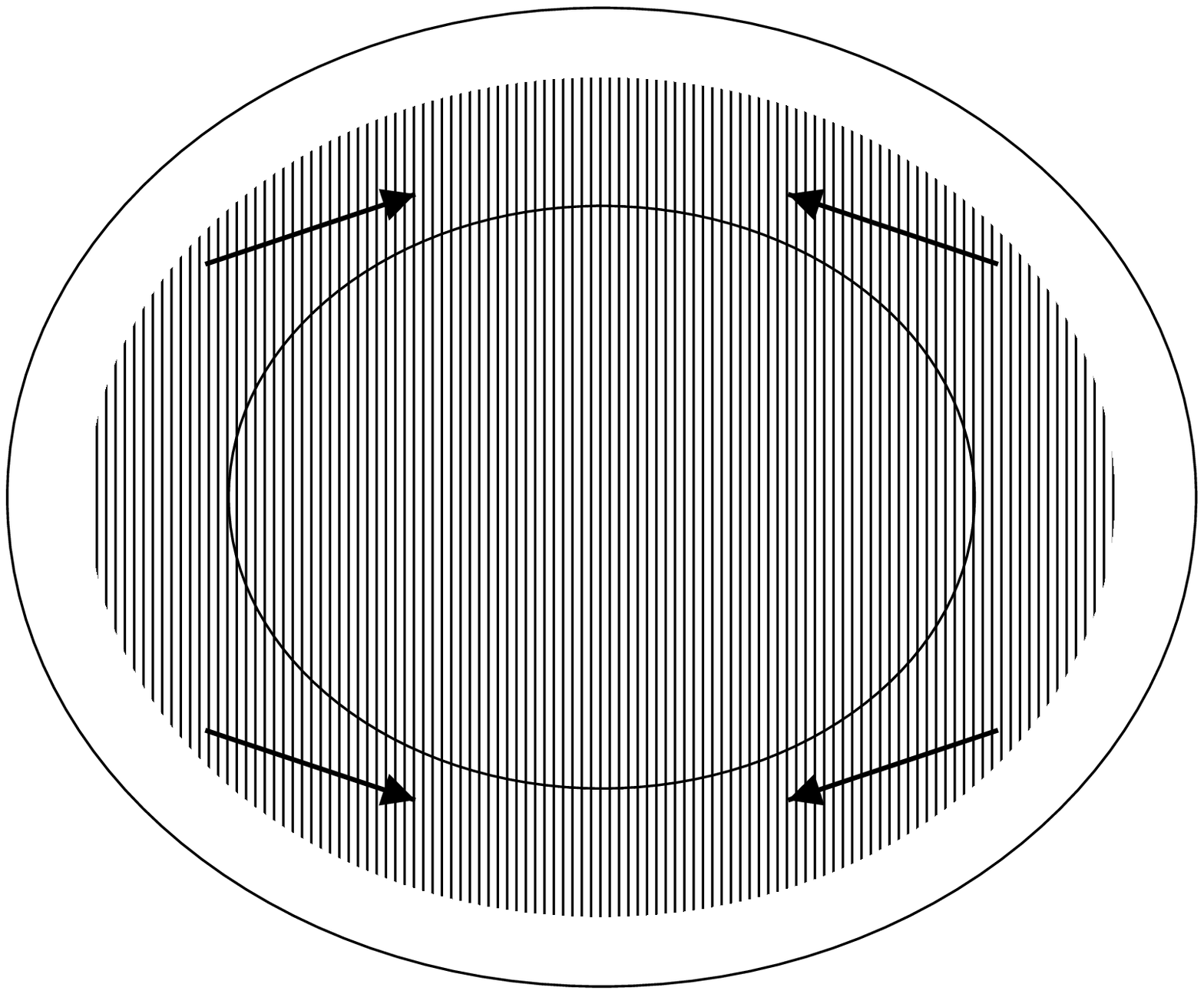, width=2.5 in} 
\caption{Qualitative sketch (using same shading conventions as before) 
indicating direction of force expected  to act on (magnetically slowed 
down) crust, even if vortex pinning is ineffective, due to centrifugal 
buoyancy mechanism when the (interpenetrating) neutron superfluid 
retains a higher rotation rate.  
\label{figC3}} 
\end{figure}

Unlike the original single constituent mechanism~\cite{R69,BPPR69,S70,BP71}
based on the loss of moment of inertia due to decrease in oblateness, and
unlike the versions~\cite{LE91,LE96,SC99} of the two-constituent theory 
that attribute the glitches to discontinuous vortex slippage, but like 
the Ruderman version~\cite{R76,R91} (that applies when the pinning is 
too strong to be broken) the newly proposed two-constituent 
mechanism~\cite{CLS00} (that applies when the pinning is too weak to be 
effective) will also entail a substantial rate of long term drift of 
plates of crust material. However this centrifugal buoyancy deficit 
mechanism differs from Ruderman's pinning driven mechanism in a manner that 
may be experimentally observable, since it is expected to produce plate 
drift in just the opposite direction, meaning that of decreasing colatitude 
$\theta$ for an isolated spinning down pulsar (see Figure~\ref{figC3}), 
entailing transfusion of matter into the crust constituent near the 
equator, and out of it nearer the poles where $\theta$ is small. In this 
mechanism (unlike Ruderman's) the angular momentum of a crust plate will 
not be significantly changed when its colatitude undergoes a displacement 
$\delta\theta$ during a glitch, so its change of rotation frequency can 
be estimated as being given roughly by $\delta \Omega/2\Omega \approx 
-\delta\theta$ where $\delta\Omega$ is the glitch amplitude that is 
actually observed, and that partially cancels the preceeding interglitch 
variation $\Delta\Omega$. The change observed in the long run is the sum 
over the glitches of the combination $\Delta\Omega+\delta\Omega$, which 
will be the same as the sum of the hidden changes $\delta_{\rm n}$  
(since the interglitch variation $\Delta_{\rm n}$ of the relevant 
superfluid part is assumed to be negligible). Since
$\delta\Omega_{\rm n}=-\big(I_{\rm c}/I_{\rm n}\big)\delta\Omega$,
by (\ref{016}), it can be seen to follow that the long run average 
of the angular drift rate will be given by
\begin{equation}\label{041} \langle\dot\theta\rangle\approx
{I_{\rm n}\over I_{\rm c}}\,{\langle\dot\Omega\rangle\over 2\Omega}
\, ,\end{equation}
which has the opposite sign to what is given by the Ruderman formula
(\ref{040}), but can be comparable in magnitude since $I_{\rm n}$
can be comparable with $I_{\rm c}$. Indeed, in a case for which
the moment of inertia $I_{\rm n}$ of the hidden part is large compared 
with the crust contribution, the magnitude given by the new formula 
(\ref{041}) would be correspondingly larger than in the previous case, with 
the implication that the crust material would be entirely recycled several 
times during the spin down lifetime of the star, while this lifetime itself
would presumably be considerably prolonged because the magnetic dipole
axis would be dragged towards the pole, thereby decreasing the pulsar
radiation rate.
 
For the purpose of observational discrimination between cases involving
strong~\cite{R76,R91}, moderate~\cite{LE91,LE96}, or very weak~\cite{CLS00}, 
coupling the relevant directly measurable parameter is what is 
known~\cite{ST83} as the (long term) braking index, 
\begin{equation}\label{043} {\rm n}={\langle\Omega\rangle
\langle\ddot\Omega\rangle\over\langle\dot\Omega\rangle^2}\, ,\end{equation}
and more particularly the braking deficit
\begin{equation}\label{044} \epsilon=3-{\rm n}\,\end{equation}
between the value that is observed and the value, ${\rm n}=3$, that is 
predicted~\cite{P67} for a simple, rigid, non aligned magnetic dipole 
model, and also for more sophisticated pulsar emission models including 
allowance~\cite{GJ69} for outflow of charged particles. If it is assumed 
that particle outflow and changes of moment of inertia can be neglected, 
then according to the simple dipole model~\cite{P67} the relative spin 
down rate $\dot\Omega/\Omega$ is just proportional to $(\Omega\, {\rm sin}
\,\alpha)^2$ where $\alpha$ is the dipole misalignment angle, i.e. the 
colatitude of the magnetic pole, so for this case the difference 
(\ref{044}) can be immediately evaluated as 
\begin{equation}\label{045} \epsilon\simeq - {2\Omega\,
\langle\dot\alpha\rangle\over {\rm tan} \,\alpha \, 
\langle\dot\Omega\rangle}\, .\end{equation}
When it gets near the extreme polar or equatorial values, $\alpha\simeq 0$, 
or $\alpha\simeq\pi/2$ , the evolution of the misalignment angle will of 
course have to come to a halt, $\langle\dot\alpha\rangle\simeq 0$, but
in the intermediate range, i.e. for ${\rm \tan}\,\alpha \approx 1$ one 
would expect the misalignment angle to move with the crustal
drift, $\langle\dot\alpha\rangle\simeq\langle\dot\theta\rangle$.
Subject to this assumption, the Ruderman formula (\ref{040}) for
the strong pinning model leads to the positive estimate
\begin{equation}\label{048} \epsilon\approx {\rm cot}\,\alpha
\, ,\end{equation}
while the formula (\ref{041}) for the model with negligible
pinning gives the negative estimate
\begin{equation}\label{049} \epsilon\approx -{I_{\rm n}\over
I_{\rm c}} {\rm cot}\,\alpha \, .\end{equation}
The effects of variation of the moment of inertia~\cite{GPW97}
should not substantially effect the validity of these estimates
except for new born pulsars with extremely rapid rotation, but
electromagnetic effects of various kinds~\cite{BGI93,Mel97} 
(including the obvious possibility of magnetic field decay) are
more likely to give significant, typically positive,
contributons to $\epsilon$. (In glitch free scenarios, 
Sedrakian and Cordes~\cite{CS98} have pointed out that
differential rotation may bring $\epsilon$ down to negative values
for periods of limited duration, but this sort of effect can be
expected to cancel out in a long term average over many gliches).

If these estimates are indeed applicable, then in the case of the Vela
pulsar the observed value~\cite{LPSC96}, namely $\epsilon\simeq 3/2$, 
can be plausibly construed as evidence favouring the Ruderman model, 
with a moderate misalignement angle of the order of 40 degrees. A less 
clear cut case example is that  of the Crab, for which the observed 
value~\cite{LPS88}, namely $\epsilon\simeq 1/2$, is also positive but  
considerably smaller, which suggests that this may be another instance 
to which Ruderman model applies, though with a relatively high 
misalignment angle. However in view of the above mentionned 
likelihood~\cite{BGI93,Mel97} of other of other positive contributions 
to $\epsilon$, this evidence is too inconclusive to exclude the 
possibility that the Crab glitches may, after all, be attributable 
a slippage mechanism~\cite{LE91,LE96,SC99} of the first category, or 
even to the original simple oblateness mechanism subject to (\ref{015}).

What transpires from all the work that has been rather rapidly surveyed 
in the preceeding sections is that the available theory of the 
internal structure of neutron stars seems to provide all the elements 
needed to account for the accumulated pulsar frequency data within the 
framework of scenarios in which superfluidity and differential 
rotation commonly play an essential role. However we are still a long 
way short of reaching any concensus about the detailed modelling of 
the many different kinds of behaviour observed in particular cases such 
as the Crab and Vela pulsars. Before any definitive understanding can 
be reached it will be necessary to carry out much more work on the 
technicalities of basic physical processes, particularly those involving 
electromagnetic effects, which were barely mentioned in the preceeding 
overview, but that are extremely important for the detailed estimation of 
important quantities such as the drag to lift ratio, $c_{\rm r}$.

The remainder of this article will be concerned with just one
of the many technical problems that needs to be dealt with
before a satisfactorily complete understanding can be achieved.
This is the problem of developing an appropriately relativistic
treatment of the superfluidity that has been seen to play such
an essential role in accounting for cases such as that of Vela,
and even the less extreme case of the Crab.

\section{Essentials of relativistic superfluid mechanics}

\subsection{Motivation and background.}

In calculations of global quantities such as the mass and radius of a 
neutron star with a given baryon number, it has been known since before 
the earliest pulsar observations in 1968 that a fully relativistic 
treatment is indispensible for even a minimally acceptable level of 
accuracy. It is fortunate that quantities such as this~\cite{Glenden97} 
can be obtained within the framework of an exactly spherical perfect 
fluid description for which a fully relativistic treatment is easily 
applicable and has always been used. This contrasts with what has been
done about secondary effects such as precession~\cite{S77,SWC99,Mel00}, 
involving mechanisms such as elasticity  (for which a relativistic 
treatment has long been available~\cite{CQ75a,CQ75b} but is difficult 
to apply) and superfluidity (for which the relevant macroscopic 
treatment~\cite{CarterLanglois95b,Carter00a,LSC98} is 
relatively new). Even in the relatively tractable context of stationary 
axisymmetric problems, or in contexts involving the (intrinsically
relativistic) effect of gravitational radiation, superfluidity and 
superconductivity have nearly always been dealt with using a  
non-relativistic Newtonian, even in relatively recent 
work~\cite{LindblomMendel94,Lee95,SW00}

There are two essentially different reasons why it is worthwhile to 
try to do better. One is of course that a fully relativistic treatment 
should in principle be more accurate, and will no doubt become 
necessary for this purpose sooner or later. However in the short run 
this is not always what is most important, since the errors inherent 
in the use of a purely Newtonian treatment range typically from a few 
per cent to a few tens of percent which is not very significant 
compared with the order of unity (or worse) uncertainties about many 
of the physical quantities involved. The other kind of reason  (which 
some readers may find surprising) is that for many purposes the use 
of a relativistic treatment is actually easier. In many cases the 
advantage of a relativistic treatment is due to the fact that the 
Lorentz group is in the technical sense semi-simple, whereas the 
Galilei group unfortunately is not. However, whether or not it is 
intrinsically simpler, the use of a relativistic treatment will 
usually be more convenient in practice whenever one wishes to use
the commonly appropriate strategy~\cite{CGPB99,CLL99,AC00} of working 
in terms of perturbations of the available spherically symmetric 
perfect fluid neutron star models. This is just because (as remarked 
above) all the best of these models (the only ones that are commonly 
taken seriously) are already formulated exclusively in a (general) 
relativisic framework. The same consideration applies to the perturbations
of the relativistic axisymmetric rapidly rotating star models that have
recently~\cite{SF98,BFG98,MSB99,GHLPBM99,AJKS00} been a subject of rapid 
development, particularly in relation to the question of bar mode 
instabilities that may be significant as a source of gravitational 
radiation.

The treatment provided here will be limited to the case of scalar (spin 0) 
models such as are appropriate for the experimentally familiar example of 
helium - 4 (though not~\cite{Volovik92} helium - 3) and also for the 
mesoscopic description of the neutron fluid that (as discussed above) 
is predicted to interpenetrate the ions in the lower crust of a 
neutron star, and that is believed~\cite{Sauls88} to condense as a 
superfluid in which the relevent bosons are scalar Cooper type pairs 
of neutrons. For the mesoscopic (intervortex) treatment of the mixed 
proton neutron superfluid below the crust a scalar description has 
commonly been employed~\cite{VS81,MenLin91} in a Newtonian treatment, and 
an analogous relativistic description~\cite{CarterLanglois98,Carter00b} 
has recently been made available. However for an exact description of such 
a mixed proton neutron superfluid, in which it is predicted~\cite{Sauls88}
that the neutrons pair up as bosons of spin 1, it would be necessary to use 
a more elaborate treatment that has yet to be developed).               

What is actually needed for the analysis of large scale effects (such as
were considered in the preceeding sections) is not just a mesoscopic
treatment of the superfluid on scales small compared with the spacing 
between interpenetrating ionic nuclei and the vortices where the 
irrotationality condition breaks down, but a macroscopic average over 
much larger scales. An appropriate macroscopic theory of the kind that 
is needed has recently been developed~\cite{LSC98} and is described 
in the final subsections of this article. The treatment presented here 
differs from the relativistic 
analogue~\cite{CarterLanglois95b,Carter00a} of the earlier non
relativistic description~\cite{BK61} of the averaged effect of vortices 
in neglecting the small anisotropy due to their effective tension, but 
instead it includes allowance for what in the long run is likely to be a 
more important effect, namely the ``transfusion'' of matter (for the 
reasons discussed in the previous subsections) between the superfluid 
constituent and the normal background, which is to be interpreted
as representing not just thermal excitations (as in ordinary
liquid helium - 4) but the entire crust component. A more elaborate 
treatment would include allowance for anisotropy of the crust constituent 
which, as noted above, will be cold enough (except very near the surface) 
to behave as an elastic solid: the way to do this has been indicated 
elsewhere~\cite{C89}, but such a mixed fluid solid description 
has not yet been developed in detail, and will not be dealt with in the 
introductory treatment provided here. 

As a preliminary to the construction of model~\cite{LSC98} that is
actually needed for the macroscopic treatment of neutron star matter, 
this presentation starts by recapitulating the long well known 
essentials of the relativistic version of the single constituent kind 
of superfluid model that is appropriate for the description of Helium 4 
at zero temperature, and of of the more recently developed 
generalisation~\cite{LebedevKhalat82,CarterKhalat92,CarterLanglois95a} 
to a two constituent model (of the kind whose non-relativistic 
analogue was originally developped by Landau) in which the second 
constituent represents a gas of phonon excitations.

\subsection{Single constituent perfect fluid models}
\label{Sect 2}

Before getting into the specific technicalities of superfluidity, 
it is worthwhile to start by  recapitulating the relevant
properties of ordinary barotropic fluid models, which includes
the category of single constituent (scalar) superfluid models
(representing the zero temperature limit of Landau type
2 constituent models) as the special case in which the vorticity 
is zero. The vorticity, in this context, is to be interpreted as the 
meaning the exterior derivative of the relevant momentum covector which 
will be formally defined below, so the vanishing of the vorticity is the 
condition for this momentum to be the gradient of a scalar potential, 
which in the superfluid case is to be understood to be proportional to the 
phase angle of an underlying bosonic quantum condensate. The qualification 
that this (zero temperature limit) model is barotropic simply means that 
there is only one independent state function such as the conserved (e.g. 
baryon) number density $n$ or the mass density $\rho$ (which are 
proportional in Newtonian theory but non-linearly related in relativistic 
theory) on which all the other state functions, such as the pressure $P$ 
are dependent. The equation of state giving $P$ as a function of $\rho$
will also determine a corresponding speed $c_{_{\rm I}}$ say, of ordinary 
``first'' sound, that will be given by the familiar formula  
\begin{equation}c_{_{\rm I}}^{\,2}=dP/d\rho\, ,\label{0a}\end{equation} 
and that must be subluminal, $c_{_{\rm I}}^{\,2}\leq c^2$ (where 
$c$ is the speed of light) in order for the usual causality requirement to  
be respected. 
 
Before proceeding it is desirable to recall the essential elements 
of the relativistic kinematics and dynamics that will be required. This 
is particularly necessary in view of the regretable tradition in 
non-relativistic fluid theory -- and particularly in non relativistic 
superfluid theory -- of obscuring the essential distinction between 
velocity (which formally belongs in a tangent bundle) and momentum (which 
formally belongs in a cotangent bundle) despite the fact that the 
distinction is generally respected in other branches of non-relativistic 
condensed matter theory, such as solid state physics, where the possibility 
of non-alignment between the 3-velocity $v^a$, and the effective 3-momentum 
$p_a$ of an electron travelling in a metallic lattice is well known. 
Failure to distinguish between contravariant entities (with ``upstairs''
indices) such as the velocity $v^a$ and covariant entities (with 
``downstairs'' indices) such as the momentum $p_a$ is something that 
one can get away with in a non-relativistic treatment only at a price that 
includes restriction to strictly Cartesian (rather than for example 
cylindrical or comoving) coordinates.  
 
In a relativistic treatment, even using coordinates $x^\mu\leftrightarrow 
\{t,x^a\}$ of Minkowski type, with a flat spacetime metric $g_{\mu\nu}$ 
whose components are of the fixed standard form diag$\{-c^2,\,1,\,1,\,1\}$, 
the necessity of distinguishing between raised and lowered indices is 
inescapable. Thus for a trajectory parametrised by proper time $\tau$, the 
correponding unit tangent vector 
\begin{equation}u^\mu={d x^\mu\over d\tau}\label{1}\end{equation}
is automatically, by construction, a contravariant vector: its space 
components, $u^a=\gamma v^a$ with $\gamma=(1-v^2/c^2)^{-1/2}$ will be 
unaffected by the index lowering operation $u^\mu\mapsto 
u_\mu=g_{\mu\nu}u^\nu$, but its time component $u^{_0}=dt/d\tau=\gamma$ 
will differ in sign from the corresponding component $u_{0}=-\gamma c^2$ 
of the associated covector $u_\mu$. On the other hand the 3-momentum $p_a$ 
and energy $E$ determine a 4-momentum covector $\mu_\nu$ with components 
$\pi_a=p_a$, $\mu_{_0}=-E$ that are intrinsically covariant. The covariant 
nature of the momentum can be seen from the way it is introduced by the 
defining equation, 
\begin{equation}\mu_\nu={\partial L\over \partial u^\nu}
\, ,\label{2a}\end{equation} 
in terms of the relevant position and velocity dependent Lagrangian 
function $L$, from which the corresponding equation of motion is obtained 
in the well known form 
\begin{equation}{d\mu_\nu\over d\tau}={\partial L\over\partial x^\nu}
\, .\label{2b}\end{equation}

In the case of a free particle trajectory, and more generally for fluid 
flow trajectories in a model of the simple barotropic kind that is 
relevant in the zero temperature limit, the Lagrangian function will have 
the familiar  standard form  
\begin{equation}L={_1\over^2}\mu g_{\mu\nu}u^\mu u^\nu-{_1\over^2}\mu c^2 
\, ,\label{3}\end{equation} 
in which (unlike what is needed for more complicated chemically 
inhomogeneous models\cite{Carter79,C89}) it is the same scalar spacetime 
field $\mu$ that plays the role of mass in the first term and that provides 
the potential energy contribution in the second term. The momentum will 
thus be given by the simple proportionality relation 
\begin{equation}\mu_\nu=\mu u_\nu\, ,\label{3a}\end{equation}
so that one obtains the expressions $E=\gamma\mu c^2$, $p_a=\mu\gamma v_a$, 
in which the field $\mu$ is interpretable as the relevant effective mass. 
 
In the case of a free particle model, the effective mass $\mu$ will of 
course just be a constant, $\mu=m$. This means that if, as we have been 
supposing so far, the metric $g_{\mu\nu}$ is that of flat Minkowski type, 
the resulting free particle trajectories will be obtainable trivially as 
straight lines. However the covarant form of the equations (\ref{1}) to 
(\ref{3a}) means that they will still be valid for less trivial cases for 
which, instead of being flat, the metric $g_{\mu\nu}$ is postulated to 
have a variable form in order to represent the effect of a gravitational 
field, such as that of a Kerr black hole (for which, as I showed in detail 
in a much earlier Les Houches school~\cite{Carter73}, the resulting non 
trivial geodesic equations still turn out to be exactly integrable). 
 
In the case of the simple  ``barotropic'' perfect fluid models with which 
we  shall be concerned here, the effective mass field $\mu$ will be 
generically  non-uniform. In these models the equation of state giving the 
pressure $P$ as a  function of the mass density $\rho$ can most 
conveniently be specified by first giving $\rho$ in terms of the 
corresponding conserved number density $n$ by an expression that will be 
decomposible in the form 
\begin{equation}\rho =mn+{\epsilon\over c^2}\, ,\label{6}\end{equation}
in which $m$ is a fixed ``rest mass'' characterising the kind of particle 
(e.g. a Cooper type neutron pair) under consideration, while $\epsilon$ 
represents an extra compression energy contribution. The pressure will 
then  be obtainable using the well known formula 
\begin{equation}P=(n\mu-\rho) c^2\, ,\label{7}\end{equation}
in which the effective dynamical mass is defined by 
\begin{equation}\mu={d\rho\over dn}=m+{1\over c^2}{d\epsilon\over dn}
\, .\label{8}\end{equation}
It is this same quantity $\mu$ (sometimes known as the ``specific 
enthalpy'') that is to be taken as the effective mass function appearing 
in the specification (\ref{3}) of the relevant Lagrangian function (on 
what is formally identifiable as the tangent bundle of the spacetime 
manifold). 
 
When one is dealing not just with a single particle trajectory but a 
spacefilling fluid flow, it is possible and for many purposes desirable 
to convert the Lagrangian dynamical equation (\ref{2b}) from particle 
evolution equation to equivalent field evolution 
equations~\cite{Carter79,C89}. Since the momentum covector $\mu_\nu$ will 
be obtained as a field over spacetime, it will have a well defined 
gradient tensor $\nabla_{\!\rho} \mu_\nu$ that can be used to rewrite the 
right hand side of (\ref{2b}) in the form $d\mu_\nu/d\tau$ $
=u^\rho\nabla_{\!\rho}\mu_\nu$. Since the value of the Lagrangian will 
also be obtained as a scalar spacetime field $L$, it will also have a 
well defined gradient which will evidently be given by an expression of 
the form 
$$\nabla_{\!\nu} L={\partial L\over\partial x^\nu}+{\partial L\over\partial 
\mu_\rho}\nabla_{\!\nu}\mu_\rho\, . $$ 
We can thereby rewrite the Lagrangian dynamical equation (\ref{2b}) as a 
field equation of the form 
\begin{equation}u^\rho\nabla_{\!\rho}\mu_\nu+\mu_\rho\nabla_{\!\nu}u^\rho
=\nabla_{\!\nu}L \, .\label{12}\end{equation}
 
An alternative approach is of course to start from the corresponding 
Hamiltonian function, as obtained in terms of the position and momentum 
variables (so that formally it should be considered as a function on the 
spacetime cotangent bundle) via the Legendre transformation 
\begin{equation}H=\mu_\nu u^\nu-L\, .\label{13}\end{equation}
In this approach the velocity vector is recovered using the formula 
\begin{equation}{dx^\mu\over d\tau}={\partial H\over \partial \mu_\nu}\, , 
\end{equation}
and the associated dynamical equation takes the form 
\begin{equation}{d\mu_\nu\over d\tau}=-{\partial H\over\partial x^\nu}
\, .\label{14}\end{equation}
The consideration that we are concerned not just with a single trajectory 
but with a spacefilling fluid means that, as in the case of the preceeding 
Lagrangian equations, so in a similar way this familiar Hamiltonian 
dynamical equation can also be converted to a field equation which takes 
the form  
\begin{equation}2 u^\rho\nabla_{\![\rho}\mu_{\nu]}=-\nabla_{\!\nu} H
\, ,\label{15}\end{equation} 
with the usual convention that square brackets are used to indicate index 
antisymmetrisation. On contraction with $u^\nu$ the left hand side will 
evidently go out, leaving the condition 
\begin{equation}u^\nu\nabla_{\!\nu}H=0 \, ,\label{16}\end{equation}
expressing the conservation of the value of the Hamiltonian allong the 
flow lines. 
 
The actual form of the Hamiltonian function that is obtained from the 
simple barotropic kind of Lagrangian function (\ref{3}) with which 
we are concerned will evidently be given by 
\begin{equation}H={1\over 2\mu} g^{\nu\rho}\mu_\nu \mu_\rho+
{\mu c^2\over 2} \, ,\label{17}\end{equation}
in which it is again the same scalar spacetime field $\mu$ that plays the 
role of mass in the first term and that provides potential energy 
contribution in the second term.  
 
In order to ensure the proper time normalisation for the parameter 
$\tau$ the equations of motion (in whichever of the four equivalent 
forms (\ref{2b}), (\ref{12}), (\ref{14}), (\ref{15}) may be preferred) 
are to be solved subject to the constraint that the numerical value of 
the Hamiltonian should vanish, 
\begin{equation}H=0\, ,\label{18}\end{equation}
initially , and hence also by (\ref{16}) at all other times. This is 
evidently equivalent to imposing the standard normalisation condition 
\begin{equation} u^\mu u_\mu=-c^2\, ,\label{19}\end{equation}
on the velocity four vector. In more general ``non-barotropic'' systems,  
such as are needed for some purposes, the Hamiltonian may be constrained 
in  a non uniform manner~\cite{Carter79,C89} so that the term on the right  
of (\ref{15}) will be non zero, but in the simpler systems that suffice 
for  our present purpose the restraint (\ref{18}) ensures that this final 
term will drop out, leaving a Hamiltonian equation that takes the very 
elegant and  convenient form 
\begin{equation}u^\nu w_{\nu\rho}=0\, .\label{20}\end{equation}
in terms of the 
relativistic vorticity tensor that is defined as the antisymmetrised 
(``exterior'')  derivative of the momentum covector, i.e. 
\begin{equation}w_{\nu\rho}=2\nabla_{[\!\nu}\mu_{\!\rho]}
\, .\label{25}\end{equation}
 
It is an evident consequence (and, as discussed in greated detail in the  
above cited Les Houches notes~\cite{Carter00a}, would still be true even 
if (\ref{18}) were not satisfied) that if $w_{\mu\nu}$ is zero initially 
it will remain zero throughout the flow, which in this case will 
be describable as ``irrotational''. 
 
In cases for which the vorticity is non-zero, the ``barotropic'' dynamical  
equation (\ref{20}) is interpretable as requiring  the flow vector $u^\mu$  
to be a zero eigenvalue eigenvector of the vorticity tensor $w_{\mu\nu}$, 
which is evidently possible only if its determinant vanishes, a requirement 
that is expressible as the degeneracy condition 
\begin{equation} w_{\mu[\nu}w_{\rho\sigma]}=0
\, .\label{26}\end{equation}
Since the the possibility of it having matrix rank 4 is thus excluded, 
it follows that unless it actually vanishes the vorticity tensor must  
have rank 2 (since an antisymmetric matrix can never have odd integer  
rank). This means that the flow vector $u^\mu$ is just a particular  
case within a whole 2-dimensional tangent subspace of zero eigenvalue  
vorticity eigenvectors, which (by a well known theorem of differential 
form  theory)  will mesh together to form well defined vorticity 
2-surfaces  as a consequence of the Poincar\'e closure property,  
\begin{equation}\nabla_{[\mu}w_{\nu\rho]}=0\, ,\label{40}\end{equation}
that follows from the definition (\ref{25}). 
  
Although it has long been well known to specialists~\cite{Lichnerowicz67},  
the simple form (\ref{20}) of what is interpretable just as the  
relativistic version of the classical  Euler equation is still not as  
widely familiar as it ought to be, perhaps because its Hamiltonian  
interpretation was not recognised until relatively  
recently~\cite{Carter79,C89}. It does not constitute by itself  
the complete set of dynamical equations of motion for the perfect fluid,  
but must be supplemented by a particle conservation equation of the usual 
form for the particle number current 
\begin{equation}n^\nu=n u^\nu\, ,\label{47}\end{equation}
which must of course satisfy the condition 
\begin{equation}\nabla_{\!\nu} n^\nu=0\, .\label{48}\end{equation}
 
A much more widely known, but for computational purposes (particularly in 
curved spacetime) less useful form of the perfect fluid dynamical 
equations is to express them in terms of the stress momentum energy density 
tensor, which is given in terms of the mass density $\rho$ and the pressure 
$P$ by 
\begin{equation}T^{\mu\nu}=(\rho+{P\over c^2})u^\mu u^\nu +P g^{\mu\nu}
\, ,\label{46}\end{equation}
and which must satisfy a so called conservation law of the standard form 
\begin{equation}\nabla_{\!\nu}T^{\mu\nu}=0
\, .\label{45}\end{equation}
Although it is conveniently succinct, a disadvantage of this traditional 
formulation is that it is directly interpretable as a law of conservation 
of momentum and energy in the strict sense only in the case of a flat 
(Minkowski type) spacetime, but not in a curved background such as that 
of a neutron star. The possibility in the barotropic case (i.e. when $P$ 
is a function only of $\rho$) of decomposing the combined set of 
dynamical equations (\ref{46}) as the combination of the convergence 
condition (\ref{48}) (obtained by contracting (\ref{45}) with $u_\mu$) and 
the relativistic Euler equation (\ref{20}), which can be written out more 
explicitly as 
\begin{equation}n^\nu\nabla_{\![\nu}\mu_{\rho]}=0
\, ,\label{49}\end{equation}
 has the advantage that these are interpretable as genuine 
conservation laws -- for particle number flux and vorticity respectively 
-- even in an arbitrarily curved spacetime background.   
 
\section{Single constituent superfluid models} 
\label{Sect 3} 
 
The simplest superfluid models, namely those pertaining to the zero 
temperature limit, are just ordinary  perfect fluid models subject to the 
restraint of irrotationality, with a momentum covector given as the 
gradient  
\begin{equation}\mu_\nu=\nabla_\nu S\, ,\label{43}\end{equation} 
of a scalar field $S$. This scalar field  is to interpreted as being 
proportional to the angle of the mesoscopic phase factor, 
${\rm e}^{i\phi}$ say, of an  underlying scalar bosonic condensate, in 
which the phase angle $\phi$ is given according to the usual 
correspondence principle by 
\begin{equation}\phi=S/\hbar\, .\label{43a}\end{equation}
In the most familiar application the bosons are Helium-4 atoms, while
between the ions of a neutron star crust below the neutron drip
transition they will be Cooper  type  neutron pairs. (However a less 
simple description is not sufficient for the spin 1 neutron pairs below 
the base of the crust, nor in the even more complicated, though 
experimentally accessible, case~\cite{Volovik92} of Helium-3, for which 
a relativistic description is still not available).

In a multiconnected configuration of a classical irrotational fluid the  
Jacobi action field  $S$ obtained from (\ref{43}) might have an arbitrary 
periodicity, but in a superfluid there will be a U(1) periodicity 
quantisation requirement that the periodicity of the phase angle $\phi$ 
should be a multiple of $2\pi$, and thus that the periodicity of the Jacobi 
action $S$ should be a multiple of $2\pi \hbar$. The simplest configuration 
for any such superfluid is a uniform stationary state in a flat Minkowski 
background, for which the phase will have the standard plane wave form  
\begin{equation}S/\hbar =k_a x^a-\omega t\, ,\label{44}\end{equation}
from which one obtains the correspondence $\mu_\nu\leftrightarrow 
\{-\hbar \omega, \hbar k_a\}$, which means that the effective 
energy per particle will be given by $E=\gamma\mu c^2=\hbar\omega$ 
and that the 3-momentum will be given by $p_a=\mu\gamma v_a=\hbar k_a$. 
 
It is to be remarked that for ordinary timelike superfluid particle 
trajectories the corresponding phase speed $\omega/k$ of the wave 
characterised by (\ref{44}) will always be superluminal -- a fact of which 
people working with laboratory Helium-4 tend to be blissfully unaware, and 
can usually safely ignore, since what matters for most practical purposes 
is not the phase speed but the group velocity of perturbation wave packets.  
 
In the irrotational case characterised by (\ref{43}) the Euler equation 
(\ref{20}) is satisfied automatically, so the only dynamical equation 
that remains is (\ref{48}). When the phase scalar is subject to a 
small perturbation, $\delta\phi=\varphi$ say, it can be seen that the 
corresponding perturbation of the conservation law (\ref{48}) provides  
a wave equation of the form 
\begin{equation}\widetilde{\Square}\varphi =0
\, ,\label{30}\end{equation}
in which $\widetilde{\Square}$ is a modified Dalembertian type operator 
that is constructed from an appropriately modified space-time metric 
tensor $\widetilde g{^{\mu\nu}}$ in the same way that the ordinary 
Dalembertian operator $\Square\equiv\nabla^\mu\nabla_{\!\mu}$ is 
constructed from the ordinary spacetime metric tensor $g^{\mu\nu}$. The 
appropriately modified spacetime metric, namely the relativistic version 
of what is known in the  context of Newtonian fluid\cite{Unruh81,Unruh95} 
and  superfluid~\cite{Visser00} mechanics as the Unruh metric,  
can be read out in terms of the light speed $c$ and the (first) sound 
speed  $c_{_{\rm I}}$ given by (\ref{0a}) as 
\begin{equation}\widetilde g{^{\mu\nu}}= {\mu\over n}\big(
c_{_{\rm I}}\gamma^{\mu\nu} 
-c_{_{\rm I}}^{\ -1}u^\mu u^\nu\big)\, ,\label{31}\end{equation}
where $\gamma^{\mu\nu}$ is the spacially projected (positive indefinite) 
part of the ordinary space time metric, as defined by 
\begin{equation}\gamma^{\mu\nu}=g^{\mu\nu}+c^{-2} u^\mu u^\nu
\, .\label{32}\end{equation}
The quantum excitations of the linearised perturbation field $\varphi$ 
governed by (\ref{30}) are what are known as phonons. For such excitations 
the phase speed and the group velocity are the same, both being given with 
respect to the unperturbed background by the ordinary (``first'') 
soundspeed, $c_{_{\rm I}}$, as given by (\ref{0a}), which will of course 
be subluminal. Phonons do nevertheless have a tachyonic aspect of their 
own, because the fact that their phase speed is subluminal automatically 
implies that they have a 4-momentum covector that is spacelike, in contrast 
with that of a ordinary fluid or superfluid particle which is timelike. 
This means that whereas the effective energy $E$ of an ordinary fluid or 
superfluid particle is always positive, the effective energy $E$ of a 
phonon may be positive or negative, depending on whether the frame of 
reference with respect to which it is measured is moving subsonically or 
supersonically. The well known implication is that if the superfluid is in 
contact with a supersonically moving boundary there will inevitably be an 
instability giving rise to dissipative phonon creation.  
 
Given a dynamical system, one of the first things any physicist is 
inclined  to ask is whether it is derivable from a Lagrangian type 
variation principle.  We have already seen in the previous sections that 
(\ref{20}) by itself is obtainable from Lagrangian equations of motion 
for the individual trajectories, which are of course obtainable from a one 
dimensional action integral of the form $\int L\, d\tau $ with $L$ as 
given by (\ref{3}). The question to be adressed now is how to obtain the 
complete set of dynamical equations (\ref{45}), including (\ref{20}) as 
well as (\ref{48}), from an action integral over the 4-dimensional 
background manifold ${\cal S}^{(4)}$  of the form 
\begin{equation}{\cal I}=\int {\cal L}\, d{\cal S}^{(4)}\, , 
\hskip 1 cm d{\cal S}^{(4)}= {\Vert g\Vert^{1/2}\over c} d^4 x\, , 
\label{50}\end{equation}
for some suitable scalar Lagrangian functional ${\cal L}$. 
 
There are several available procedures for doing this for a generic 
perfect fluid with rotation, involving radically different choices of the 
independent variables to be varied: although they are all ultimately 
equivalent ``on shell'' the ``off shell'' bundles over which the variations 
are taken differ not only in structure but even in dimension. These methods 
notably include the worldline variation procedure (the most economical from 
a dimensional point of view) developed by Taub\cite{Taub54}, and the 
Clebsch type variation proceedure developed by Schutz\cite{Schutz70}, as 
well as the more recently developed  Kalb-Ramond type 
method\cite{Carter94,Carter00a} that has been specifically designed for 
generalisation~\cite{CarterLanglois95b} to allow for 
the anisotropy arising from the averaged effect of vortex tension in the 
treatment of superfluidity at a macroscopic level. None of these various 
methods is sufficiently simple to have become widely popular. 
 
The problem is much easier to deal with if, to start off with, one 
restricts oneself to the purely irrotational case characterised by 
(\ref{43}), which is all that is needed for the description of zero 
temperature superfluidity at a mesoscopic level. In this case a very simple 
and well known procedure is available. In this procedure, the independent 
variable is taken to be just the Jacobi action $S$, or equivalently in a 
superfluid context, the phase $\varphi$ as given by (\ref{43a}), and the 
action is simply taken to be the pressure P expressed as a function of the 
effective mass $\mu$, with the latter constructed as proportional to the 
amplitude of the 4-momentum, according to the prescription 
\begin{equation}\mu^2 c^2=-\mu_\nu \mu^\nu
\, , \label{52}\end{equation}
with the 4-momentum itself given by the relation (\ref{43}) that applies 
in the irrotational case, i.e. 
\begin{equation}\mu_\nu=\hbar\nabla_\nu\phi
\, .\label{53}\end{equation}
Thus setting 
\begin{equation}{\cal L}=P\, ,\label{54}\end{equation}
and using the standard pressure variation formula 
\begin{equation}\delta P= c^2 n \delta \mu
\, ,\label{54a}\end{equation}
one sees that the required variation of the Lagrangian will be given by 
\begin{equation} \delta{\cal L}=-n^\nu\delta\mu_\nu=-\hbar 
n^\nu\nabla_{\!\nu}(\delta \phi)\, .\label{54b}\end{equation}
Demanding that the action integral (\ref{50}) be invariant with 
respect to infinitesimal variations of $\varphi=\delta\phi$ then evidently  
leads to the required conservation law (\ref{48}).  
 
It is to be noted that this variational principle can be reformulated in 
terms of an independently variable auxiliary field amplitude $\Phi$ and an 
appropriately constructed potential function $V\{\Phi\}$ as a function of 
which the action takes the desirably fashionable form 
\begin{equation}{\cal L}=-{\hbar^2\over 2}\Phi^2(\nabla_{\!\nu}\phi) 
\nabla^\nu\phi-V\{\Phi\}\, , \label{56}\end{equation}
which is interpretable as the classical limit of a generalised Landau 
Ginzburg type model. In this formulation, as discussed in greater detail 
elsewhere~\cite{Carter94,Carter00a}, the auxiliary amplitude is to be 
identified as being given by the formula   
\begin{equation}\Phi={n\over\sqrt{\rho+{P/ c^2}}} 
=\Big({n\over\mu}\Big)^{1/2}\, ,\label{62}\end{equation}
while the prescription for the corresponding potential energy density 
function is that it should be given by 
\begin{equation}V={\rho c^2-P\over 2}\, .\label{63}\end{equation}
Having evaluated $V$ as a function of $\Phi$ one can recover the 
effective mass $\mu$, number density $n$, mass density $\rho$ 
and pressure $P$ of the fluid using the formulae 
\begin{equation}\mu^2={1\over c^2\Phi}{dV\over d\Phi}\, ,\hskip 1 cm n=
\Phi^2 \mu\, , \label{63a}\end{equation}
and  
\begin{equation}\rho={_1\over^2}\,\Phi^2\mu^2+{V\over c^2}\, ,\hskip 1 cm 
P={_1\over^2}\,\Phi^2\mu^2 c^2 -V\, ,\label{63b}\end{equation}
which are derivable from (\ref{7}) and (\ref{8}). It is to be remarked 
that the covariant inverse of the generalised Unruh tensor (\ref{31}) 
is expressible in this notation as 
\begin{equation}\widetilde g^{-1}_{\ \mu\nu} =\Phi^2  \big( 
c_{_{\rm I}}^{\ -1}\gamma_{\mu\nu} 
-c_{_{\rm I}} c^{-2}u_\mu u_\nu\big)\, .\label{65} \end{equation}
 
A particularly noteworthy example is the conformally invariant special 
case~\cite{Carter00a} characterised by a potential function that is 
homogeniously quartic, $V\propto\Phi^4$, which is what is obtained for a 
radiation gas type equation of state of the familiar form $P=\rho c^2/3$, 
and for which the (first) sound speed is given by 
$c_{_{\rm I}}^{\, 2}= c^2/3$. 
 
\section{Landau type 2-constituent superfluid models.} 
\label{Sect 4} 
 
As an intermediate step between the very simple single constituent 
superfluid models described in the previous section and the more 
elaborate models needed in the context of neutron stars, the purpose of 
this section is to describe the relativistic version of the category of 
non dissipative 2-constituent superfluid that was originally developed by 
Landau for the description of ordinary superfluid Helium-4 at non-zero 
temperature. As well as the relevant conserved particle number current 
$n^\mu$ (representing the flux of Helium atoms in that particular 
application) such a model involves another independently conserved current 
vector, $s^\mu$ say, representing the flux of entropy. In the single 
constituent case characterised by the variation rule (\ref{54b}) we saw 
how  the current vector $n^\nu$ was associated with a dynamically 
conjugate covector $\mu_\nu$ that is interpretable as representing the 
effective mean 4-momentum per particle. In a similar way in a 
2-constituent model the second current vector $s^\nu$ will be analogously 
associated with its own dynamically conjugate 4-momentum covector 
$\Theta_\nu$.  
           
The earliest presentations of the generic category of non-dissipative 
2-constituent superfluid were on one hand a 
generalisation~\cite{LebedevKhalat82} of the relativistic Clebsch 
formulation~\cite{Schutz70} based on the variation of a generalised 
pressure function $\Psi$ depending on the 4-momentum covectors 
$\mu_\nu$ and $\Theta_\nu$ according to the partial differentiation rule 
\begin{equation}d\Psi=-n^\nu\, d\mu_\nu-s^\nu\, d\Theta_\nu 
\, ,\label{70}\end{equation}
and on the other hand a  generalisation~\cite{Carter85,C89} of the 
world line variational formulation due to Taub~\cite{Taub54} based on 
the variation of a master function $\Lambda$ depending on the currents 
$n^\mu$ and $s^\nu$ according to the partial differentiation rule 
\begin{equation}d\Lambda= \mu_\nu\, dn^\nu + \Theta_\nu\, ds^\nu
\, .\label{71}\end{equation}
Although they were originally developed independently these alternative 
formulations were subsequently shown to be equivalent to each other and to 
an intermediate crossbred version~\cite{CarterKhalat92} based on a 
Lagrangian density 
\begin{equation}{\cal L}=\Psi+s^\nu\Theta_\nu=\Lambda-n^\mu\mu_\nu
\, ,\label{72}\end{equation} 
depending on the particle 4-momentum covector $\mu_\nu$ and the entropy 
current $s^\nu$ according to the partial differentiation rule 
\begin{equation}d{\cal L}=\Theta_\nu\, ds^\nu-n^\nu\, d\mu_\nu
\, .\label{73}\end{equation}
 
All of these variational formulations are subject to the complication 
that the allowable field variations are not free but must be suitably 
constrained to avoid giving overdetermined field equations. Although it 
violates the symmetry between the two kinds of conserved current $n^\nu$ 
and $s^\nu$ that are involved, the crossbred formulation characterised by 
(\ref{73}) is the one that allows the simplest specification of the 
constraints required to get the appropriate dynamical equations for the 
superfluid case. In this  formulation~\cite{CarterKhalat92} the constraint 
on the particle 4-momentum  covector is simply that it should have the 
same phase gradient form (\ref{43}) as in the zero temperature limit in 
which the entropy constituent is absent,  namely 
\begin{equation}\mu_\nu=\hbar\nabla_{\!\nu}\phi
\, .\label{75}\end{equation}
 
The corresponding constraint on the current vector $s^\nu$ of the 
``normal'' constituent is the not quite so simple Taub type requirement 
that its variation should be determined by the displacement of the flow 
lines generated by an arbitrary vector field $\zeta^\nu$ say, which 
means~\cite{C89} that it must have the form 
\begin{equation}ds^\nu=\zeta^\rho\nabla_{\!\rho} s^\nu-s^\nu
\nabla_{\!\rho}\zeta^\nu +s^\nu\nabla_{\!\rho}\zeta^\nu
\, ,\label{76}\end{equation}
whose derivation is obtainable by a procedure that will be explained 
more explicitly in the next Section. Demanding invariance of the volume 
integral  of ${\cal L}$ with respect to infinitesimal local variations of 
the phase variable  $\phi$ then gives the usual particle conservation law 
in the same form (\ref{48}) as for the single constituent limit, while 
demanding invariance for an arbitrary local displacement field $\zeta$ 
gives not only the analogous entropy conservation law 
\begin{equation}\nabla_\nu s^\nu=0\, ,\label{77}\end{equation}
but also the dynamical equation, 
\begin{equation}s^\nu\nabla_{[\nu}\Theta_{\rho]}=0
\, , \label{78}\end{equation}
that governs the evolution of the thermal 4-momentum covector in a manner 
analogous to that whereby the relativistic Euler equation (\ref{49}) 
governs the evolution of the momentum covector in an ordinary perfect 
fluid. These dynamical equations entail (but unlike the single constituent 
case are not entirely contained in) an energy momentum pseudo-conservation 
law of the usual form (\ref{45}) for a stress-momentum-energy density 
tensor that can be written in the form 
\begin{equation}T^\nu_{\,\rho}=n^\nu\mu_\rho+s^\nu\Theta_\rho+
\Psi g^\nu_{\,\rho} \, ,\label{80}\end{equation}
which will in fact (although it is not obvious in this particular 
expression) be automatically symmetric, $T^{[\nu\rho]}=0$.  
         
The category of models characterised by the preceeding specifications 
for various conceivable forms of the equation of state specifying 
${\cal L}$ as a scalar function of $\mu_\nu$ and $s^\nu$ is very 
large. The use of what is interpretable~\cite{CarterKhalat92b} 
as a special subcategory therein, on the basis of a particular 
kind of separation ansatz, was proposed in early work of  
Israel~\cite{Israel81} and Dixon~\cite{Dixon82} and has been advocated 
more recently by Olsen~\cite{Olsen90}. Unfortunately however, the  
simplification provided by the Israel Dixon ansatz (effectively the 
relativistic generalisation of the obsolete Tisza-London theory that 
was superceded by that of Landau) is incompatible with the kind of  
equation of state that is needed for even a minimally realistic 
treatment of a real superfluid.

A satisfactory treatment of what goes on at temperatures high enough for 
non-linear ``roton'' type excitations to be important is not yet 
available, but in the low temperature ``cool'' regime, in which only 
linear ``phonon'' type excitations are important, it is not difficult to 
provide a straightforward analytic derivation of the kind of equation of 
state that is appropriate. Following the lines developed in a non - 
relativistic context by Landau himself~\cite{Landau} the relativistic 
version of the approriate ``cool'' equation of state has recently been 
derived~\cite{CarterLanglois95a} by considering perturbations of the 
single constituent model -- with equation of state specified as a pressure 
function, $P\{\mu\}$ -- that describes the relevant zero temperature limit. 
The result is obtained in an analytically explicit form that (despite the 
fact that it is not of the separable Israel Dixon kind) can be given a 
very simple expression in terms of what we referred to as the ``sonic'' 
metric, which is specifiable by the conformal relation 
\begin{equation}{\cal G}^{\rho\sigma}=\Phi^2 c_{_{\rm I}}^{\ -1}  
\widetilde g^{\rho\sigma}\label{82}\end{equation}
in terms the Unruh phonon metric (\ref{31}) that is associated with the 
relevant zero temperature limit state as specified by the relevant 
momentum covector $\mu_\nu$, which by (\ref{52}) determines the relevant 
value of the scalar $\mu$ and hence (via the zero temperature equation of 
state, using the formalism of Section \ref{Sect 2}) also of the relevant 
phonon speed $c_{_{\rm I}}$ and field amplitude $\Phi$. While the Unruh 
metric is more convenient for many purposes, the advantage of the 
conformal modification we have used, namely 
\begin{equation}{\cal G}^{\rho\sigma}=g^{\rho\sigma} +\big(
c^{-2}-c_{_{\rm I}}^{\ -2} \big)u^\rho u^\sigma
\, ,\label{83}\end{equation}
is that its spacially projected part agrees with that of the ordinary 
space  metric, from which it differs only in the measurement of time. 
 
The result that is obtained~\cite{CarterLanglois95a} is given by a 
Lagrangian of the form 
\begin{equation}{\cal L}= P-3\psi \label{84}\end{equation}
in which the deviation from the zero pressure limit value $P\{\mu\}$ 
is given as a function not just of the particle 4-momentum 
covector $\mu_\nu$ but also of  the entropy flux $s^\nu$ (postulated to 
be sufficiently weak to be constituted only of phonons) by the formula 
\begin{equation}\psi ={\tilde\hbar\over 3} c_{_{\rm I}}^{\ -1/3}| 
{\cal G}^{-1}_{\,\rho\sigma} s^\rho s^\sigma|^{2/3}
\, ,\label{85}\end{equation}
where $\tilde\hbar$ is identifiable to a very good approximation 
with the usual Dirac-Planck constant $\hbar$, its exact value being 
given by 
\begin{equation}\tilde \hbar = {9\over 4\pi}\Big({5\pi\over 6}\Big)^{1/3} 
\simeq 0.99\,\hbar\, .\label{86}\end{equation}
This is equivalent to taking the generalised pressure function 
to be 
\begin{equation}\Psi = P+\psi\, ,\label{87}\end{equation}
with 
\begin{equation}\psi= {c_{_{\rm I}}\over 4}\Big({3\over 4\hbar}\Big)^3 
\big({\cal G}^{\rho\sigma}\Theta_\rho\Theta_\sigma\big)^2
\, ,\label{88}\end{equation}
in which the effective thermal 4-momentum per unit of entropy 
is given (according to the partial differentiation formula 
(\ref{73}) by 
\begin{equation}\Theta_\rho= {4\tilde\hbar\over 3}|c_{_{\rm I}} 
{\cal G}^{-1}_{\,\mu\nu} s^\mu s^\nu|^{-1/3} {\cal G}^{-1}_{\,\rho\sigma} 
s^\sigma\, ,\label{89}\end{equation}
with 
\begin{equation}{\cal G}^{-1}_{\,\rho\sigma}=g_{\rho\sigma}+  
\Big( 1 - {c_{_{\rm I}}^{\ 2}\over c^2}\Big)|\mu^\nu\mu_\nu|^{-1} 
\mu_\rho\mu_\sigma\, .\label{90}\end{equation}
 
An concrete illustration, allowing the explicit evaluation of the 
relevant quantities, is provided by the polytropic case, as 
characterised by  a (single consitituent) equation of state giving 
the mass density $\rho$ as a function of the number density $n$ in 
terms of a fixed (``rest'') mas per particle $m$, a scale constant 
$\kappa$ and a fixed dimensionless index  $\gamma$  in the form 
\begin{equation}\rho=m n+\kappa n^\gamma \Leftrightarrow \mu=m+
\kappa\gamma n^{\gamma-1} \, ,\label{92}\end{equation}
which corresponds to taking the pressure to be given by 
\begin{equation}P=\kappa c^2 (\gamma-1) n^\gamma=\kappa c^2 (\gamma-1)
\Big({\mu-m \over\kappa\gamma}\Big)^{\gamma/(\gamma-1)}
\, ,\label{93}\end{equation}
while the corresponding sound speed will be given (independently 
of $\kappa$) by 
\begin{equation}c_{_{\rm I}}^{\ 2} = (\gamma-1)\Big(1-{m\over\mu}\Big) 
c^2\, . \label{94}\end{equation}
 
\section{Non-conservative model with transfusion and vortex drag} 
\label {Sect 5}

Although the Landau type of model described in the previous section has 
been found to be very effective for the description of liquid Helium-4 
under laboratory conditions, it is not of much use for direct application 
in neutron star matter because the thermal effects it allows for will in 
general be less important than other complications whose treatment will
require the use of more elaborate models whose relativistic versions are 
still at a relatively early stage of development and will not be presented 
here. The most important of these complications, whose treatment in a 
relativistic framework has been the subject of preliminary work that is 
discussed elsewhere, are due to the effect of the protons that will be 
present, either in ionic nuclei that are responsible for the elastic solid 
behavior~\cite{C89} of the crust, or as a dissolved 
superfluid~\cite{CarterLanglois98,Carter00b} at deeper levels. Another 
complication that is relevant for the macroscopic treatment of a neutron 
star is the necessity of averaging over an Abrikosov type lattice of 
quantised vortices (that must be roughly aligned with the rotation axis of 
the star) whose effective tension entails 
deviations~\cite{CarterLanglois95b,Carter00a} from perfect fluid isotropy.

Like the thermal effect discussed in the preceding section, these various 
complications can all be provisionally set aside as perturbations to be 
incorporated at a later stage in a systematic approach whose first stage 
requires the use only of a relatively crude description in which, except 
for the superfluid neutrons with baryon number current vector 
$\nN^{\,\nu}$ all the other constituents, meaning mainly protons and 
electrons, move together with the entropy as a single ``normal'' 
constituent with baryon number current $\nC^{\,\nu}$. Whereas the 
anisotropy arising from vortex tension~\cite{CarterLanglois95b,Carter00a} 
is relatively unimportant, a major role in the long term evolution of the 
star is likely to be played by the static pinning or dynamical drag forces 
exerted on the vortices by the composite ``normal'' background constituent. 
Another effect that is of importance in the long run is that of 
``transfusion'' whereby -- due to the subduction resulting from the drift 
mechanism whose effect is roughly described by (\ref{040}) or (\ref{041})
-- the superfluid neutron contribution $\nN^{\, \nu}$ to the baryon 
current may undergoes transformation (via weak beta decay type processes) 
to the ``normal'' (essentially protonic) consituent, and vice versa, so 
that only the total baryon current  
\begin{equation}\nB=\nN^{\, \nu}+\nC^{\, \nu}
\label{2.3}\end{equation}
remains locally conserved throughout: 
\begin{equation}\nabla_{\!\nu}\nB^{\,\nu}=0\, . \label{2.4}\end{equation}
 
The kind of (non-conservative) 2-constituent model needed for this purpose 
is obtainable as a generalisation of the kind of (conservative) 
2-constituent superfluid model discussed in the preceding section, 
starting from the formulation in terms  of a master function  $\Lambda$ in 
which the currents (not momenta) are taken as the independent (but not 
entirely free) variables. 
 
In a transfusive model of the type set up here, the ``normal'' constituent 
is not entirely dependent on (though it does include) entropy, so that it 
is present even at zero temperature: the primary role of this non - 
superfluid constituent is to represent the fraction of the baryonic 
material of the neutron star that is not included in the neutron 
superfluid, as well as the degenerate electron gas that will be present to 
neutralise the charge density resulting from the fact that some of these 
baryons will have the form of protons rather than neutrons. In the solid 
``crust'' layers of a neutron star the protons will be concentrated 
together with a certain fraction of the neutrons in discrete nuclear type 
ions, which at the relatively moderate temperatures that are expected to 
apply will form a solid lattice. In the upper crust the ``normal'' 
constituent consisting of the ionic lattice and the degenerate electrons 
will include everything, but in the lower crust (at densities above about 
$10^{11}$ gm/cm$^3$) the crust will be interpenetrated by an independently 
moving neutron superfluid. What we refer to as ``transfusion'' occurs when 
compression takes place so that the ionic constituent undergoes a fusion 
process whereby neutrons are released in the form of newly created 
superfluid matter, or conversely, when relaxation of the pressure allows 
excess neutrons to be reabsorbed into the ions.

A more elaborate treatment would specifically allow for the expectation 
that the protons would form an independently conducting superfluid of 
their own at very high densities, whereas they will combine with some of 
the neutrons at intermediate densities, and with all of the neutrons at 
low densities, to form discrete ions which will tend to crystalise to 
form a possibly anisotropic lattice. What matters for our present 
purpose is that regardless of its detailed  constitution, all this 
``normal'' matter will in effect be strongly self coupled\cite{ALS84} by 
short range electromagnetic interactions so that its movement will be 
describable to a very good approximation as that of a single fluid with 
a well defined 4-velocity, $\uC{^\mu}$ say, the only independent motion 
being that of the (electromagnetically neutral) neutron superfluid with 
velocity $\uN{^\mu}$ say.  The latter will specify the direction of the 
part of the baryon current,  
\begin{equation}\nN{^\mu}=\nN \uN{^\mu}
\, ,\label{2.1}\end{equation} 
carried by the neutron superfluid, while the ``normal'' matter velocity 
specifies the direction of the remaining {\it  collectively comoving} 
part,  
\begin{equation}\nC{^\mu}=\nC\uC{^\mu} 
\, ,\label{2.2}\end{equation} 
of the baryon current.  
 
At densities below the ``neutron drip'' transition at about $10^{11}$ 
gm/cm$^3$, the ``normal''  collectively comoving constituent $\nC{^\mu}$ 
will of course be identifiable with the total, $\nB{^\mu}$.  The reason 
why the remaining free neutron part $\nN{^\mu}$ -- which will always  
be present at higher densities -- is presumed to be in a state of 
superfluidity is that the relevant condensation temperature, below which 
the neutrons form bosonic condensate of Cooper type pairs is estimated 
\cite{E88} to be at least of the order of $10^9$ K, while it is expected 
that a newly formed neutron star will drop substantially below this 
temperature within a few months~\cite{Ts79}. At such comparatively low 
temperatures the corresponding entropy current $s^\mu$ say will not play 
a very important dynamical role, but for the sake of exact internal 
consistency it will be allowed for in the model set up here, in which 
it will be taken for granted that it forms part of the ``normal'' 
collectively comoving constituent so that it will have the form 
\begin{equation}s^\mu= s\uC^\mu\, .\label{2.5}\end{equation}

Under conditions of sufficiently slow convection, the transfer needs 
not involve significant dissipation, so  the process should be 
describable by a Lagrangian scalar, $\Lamb$ say, that will depend just 
on the currents introduced above, of which the independent components 
are given just by the vectors $\nC^\mu$ and $\nN^\mu$ and the scalar 
$s$. As a first approximation (whose accuracy in the various relevant  
density regimes is a subject that needs much further investigation) one  
might suppose that the Lagrangian separates in the form  $\Lamb $ 
$=-\rhoC c^2-\rhoN c^2$ in which $\rhoC$ is a mass density depending 
only on $s$ and $\nC$, while $\rhoN$ is an another energy mass 
depending only on $\nN$, but we shall not invoke such a postulate here, 
i.e. we allow for the likelihood that the properties of ``normal'' 
constituent will be affected by the presence of the superfluid 
constituent and vice versa, which means that there will be an 
{\it entrainment} effect\cite{AB76,VS81,ALS84,Sj76}, whereby for example 
the velocity of the superfluid neutron current will no longer be 
parallel to the corresponding momentum. (As an alternative to the more 
suitable term ``entrainment'' this mechanism is sometimes referred to 
in the litterature as ``drag'', which is misleading because entrainment 
is a purely conservative, entirely non-dissipative effect, whereas the 
usual kinds of drag in physics, and in particular the kind 
of drag to be discussed below, are essentially dissipative processes.) 
 
If we adopted the (gas type) description embodied in the separation 
ansatz we would have two separate variation laws which in a fixed 
background would take the form $c^2\delta\rhoC=\Theta \delta s+c^2\muC 
\delta\nC$ and $c^2\delta \rhoN = c^2\muN \delta\nN$, in which $\Theta$ 
would be interpretable as the temperature, $\muC$ would be interpretable 
as the effective mass per baryon in the ``normal'' part, and $\muN$ would 
be efective mass per neutron in the superfluid part (which would be equal 
to its analogue in the ``normal'' part, i.e. $\muN=\muC$, in the 
particular case of a state of static thermodynamic equilibrium.) 
 
In the less specialised (liquid type) description to be used here, there 
will just be a single ``conglomerated'' variation law, whose most general 
form, including allowance for a conceivable variation of the background 
metric, will be expressible (correcting one of the copying errors in the 
originally  published version~\cite{LSC98}) as 
\begin{equation}\delta\Lamb=-\Theta\delta s +\muC{_\nu} \delta\nC^\mu+  
 \muN{_\nu}\delta\nN^\nu +{_1\over^2}\big( c^{-2}\Theta s\uC^{\,\mu} 
\uC^{\,\nu}+\nC{^\mu}\muC{^\nu}  +\nN{^\mu}\muN{^\nu}\big)
\delta g_{\mu\nu}\, ,\label{2.6}\end{equation}
where $\Theta$ is to be interpreted as the temperature and where 
$\muN_\mu$ and $\muC_\mu$ are to be interpreted as the 4-momentum per 
baryon of the neutron superfluid and the ``normal'' constituent 
respectively. 
 
To obtain suitable fluid type dynamical equations from a Lagrangian 
expressed as above just in terms of the relevant currents, the variation 
of the latter must be appropriately constrained in the manner\cite{C89} 
that was originally introduced for the case of a simple perfect fluid by  
Taub\cite{Taub54}. The standard Taub procedure can be characterised as 
the requirement that the variation of the relevant current three form, 
which for the ``normal'' constitituent in the present application will be 
given in terms of the antisymetric space-time measure tensor 
$\varepsilon_{\mu\nu\rho\sigma}$ by 
\begin{equation}\CC_{\mu\nu\rho}=\varepsilon_{\mu\nu\rho\sigma}
\nC^\sigma\, ,\label{2.7}\end{equation}
should be given by Lie transportation with respect to an associated, 
freely chosen, displacement vector field $\xiC{^\mu}$ say. This ansatz 
gives the well known result  
\begin{equation}\delta \CC_{\mu\nu\rho}=\xiC^\lambda\nabla_{\!\lambda} 
\CC_{\mu\nu\rho} +3\CC_{\lambda[\mu\nu}\nabla_{\!\rho]}\xiC{^\lambda}
\, .\label{2.8}\end{equation}
Although a variation $\delta g_{\mu\nu}$ of the metric has no effect on 
the fundamental current three form, $\CC_{\mu\nu\rho}$, it will contribute 
to the variation of the corresponding vector, 
\begin{equation}\nC^\mu={1\over 3!}\varepsilon^{\mu\nu\rho\sigma}
\CC_{\nu\rho\sigma}\, ,   \label{2.9}\end{equation}
for which one obtains 
\begin{equation}\delta \nC^\mu=\xiC^\nu\nabla_{\!\nu}\nC^\mu-
\nC^\nu\nabla_{\!_\nu} \xiC^\mu +\nC^\mu\big(\nabla_{\!\nu}\xiC^\nu-
{_1\over^2} g^{\nu\rho} \delta g_{\nu\rho} \big)
\, . \label{2.10}\end{equation}  
(Application of an analogous procedure to the entropy current provides  
the variation rule (\ref{76}) that was used in Section \ref{Sect 3}).) 
In terms of the orthogonally projected metric, 
\begin{equation}\gamC{^{\mu\nu}}=g^{\mu\nu}+c^{-2}\uC^\mu \uC^\nu
\, , \label{2.11}\end{equation}
the corresponding variation of the unit flow vector will be given by 
\begin{equation}\delta \uC{^\mu}=\gamC{^\mu}_{\, \rho}\big(\xiC^\nu
\nabla_{\!\nu}\uC^\rho -\uC{^\nu}\nabla_{\!\nu}\xiC{^\rho}\big)+
{_1\over^2}c^{-2}\uC^\mu\uC^\nu \uC^\rho \delta g_{\nu\rho}
\, ,\label{2.12}\end{equation}
and the corresponding variation in the current amplitude $\nC$ will be 
\begin{equation}\delta \nC=\nabla_{\!\nu}\big(\nC\xiC^\nu\big) 
 +\nC\big(c^{-2}\uC{^\mu}\uC{^\nu}\nabla_{\!\mu}\xiC{_\nu} 
 -{_1\over^2}\gamC{^{\mu\nu}}\delta g_{\mu\nu}\big)
\, .\label{2.13}\end{equation}
Since the entropy flux is to be considered as comoving with the 
``normal'' constituent, it is subject to a variation given by the same 
displacement vector $\xiC$, which thus gives  
\begin{equation}\delta s=\nabla_{\!\nu}\big(s\xiC{^\nu}\big)+s\big(c^{-2} 
 \uC{^\mu}\uC{^\nu}\nabla_{\!\mu}\xiC{_\nu} 
 -{_1\over^2}\gamC^{\,\mu\nu}\delta g_{\mu\nu}\big)
\, .\label{2.14}\end{equation}
On the other hand for the superfluid constituent there will be an 
independent displacement vector field $\xiN^\mu $ say, 
in terms of which the analogously constructed variation will be 
\begin{equation}\delta \nN^\mu=\xiN^\nu\nabla_{\!\nu}\nN^\mu-\nN^\nu 
\nabla_{\!\nu}\xiN^\mu +\nN^\mu\big(\nabla_{\!\nu}\xiN^\nu 
-{_1\over^2}g^{\nu\rho} \delta g_{\nu\rho} \big)
\, .\label{2.15}\end{equation}  
 
The effect of this variation process on the Lagrangian density 
$\Vert g\Vert^{1/2}\Lamb$ itself can be seen to be expressible 
in the standard form 
\begin{equation}\Vert g\Vert^{-1/2}\delta\big(\Vert g\Vert^{1/2}
\Lamb\big)=  \xiC{^\nu}\fC{_\nu}+\xiN{^\nu}\fN{_\nu}+{_1\over^2}
\TE{^{\mu\nu}} \delta g_{\mu\nu}+\nabla_{\!\mu}{\cal R}^\mu
\, ,\label{2.16}\end{equation}
in which $\fC{_\nu}$ will be interpretable as the force density acting on  
the ``normal'' constituent, $\fN{_\nu}$ will be interpretable as the force  
density acting on the superfluid constituent, and $\TE{^{\mu\nu}}$ will be  
interpretable as the stress momentum energy density of the two constituent  
as a whole. By considering the trivial case in which there is no actual 
physical alteration of the system, but in which the apparent changes are  
merely due to the displacement of the reference system generated by a  
vector field $\xiN^{\nu}=\xiC^{\nu}$, in which case the apparent  
variation of the metric will be given by 
$\delta g_{\nu\nu}=2\nabla_{\![\mu}\xiC_{\nu]}$,  
it can be seen from (\ref{2.16}) that the separate forces must  
automatically satisfy an identity of the form 
\begin{equation}\fC{_{\!\nu}}+\fN{_{\!\nu}}=\fE{_{\!\nu}}
\, , \label{2.23}\end{equation}
where $\fE{_{\!\nu}}$ is the conglomerated {\it external force density} 
that is defined by 
\begin{equation}\fE{_{\!\nu}}=\nabla_{\!\mu} \TE{^\mu}_\nu 
\, .\label{2.24}\end{equation}
 
The residual current ${\cal R}^\mu$ in the divergence will be of no 
importance for our present purpose (by Green's theorem it just gives a 
surface contribution that will vanish by the variational boundary 
conditions) but it is to be noted for the record that it will have the 
form 
\begin{equation}{\cal R}^\mu=2\xiC{^{[\mu}}\uC{^{\nu]}} \big(c^{-2}\Theta 
s\uC{_\nu}+ \nC\muC_\nu\big) +2\xiN\,{^{[\mu}}\nN{^{\nu]}}\muN{_\nu}
\, .\label{2.17}\end{equation}
The conglomerated stress momentum energy density tensor can easily be 
read  out as 
\begin{equation}\TE{^\mu}_{\nu}=\Psi g^\mu_{\ \nu}+c^{-2}\Theta 
s\uC{^\mu}\uC{_\nu} +\nC{^\mu}\muC{_\nu} + \nN{^\mu}\muN{_\nu} 
\, ,\label{2.18}\end{equation}
where 
\begin{equation}\Psi=\Lamb+s\Theta-\nC{^\nu}\muC{_\nu}-\nN{^\nu}
\muN{_\nu}\ .\label{2.19}\end{equation}
(Although this expression is not manifestly symmetric, the asymmetric 
contributions will automatically cancel due to the identity 
$\muC{^{[\mu}}\nC{^{\nu]}}=-\muN{^{[\mu}}\nN{^{\nu]}}$). 
What matters most for our present purpose is the form of the respective 
force  densities: the force law (i.e. the relevant relativistic 
generalisation of Newton's ``second'' law of motion) for the ``normal'' 
constituent is found  to take the form 
\begin{equation}\fC{_\nu}= 2s^\mu\nabla_{\![\mu}\big(c^{-2}\Theta 
\uC{_{\nu]}}\big) +2\nC{^\mu}\nabla_{\![\mu}\muC_{\,\nu]}+c^{-2}\Theta  
 \uC{_\nu}\nabla_{\!\mu} s^\mu +\muC_{\,\nu}\nabla_{\!\mu} \nC^\mu
\, ,\label{2.20}\end{equation}
while the force law for the superfluid component is found to take the 
simpler  form 
\begin{equation}\fN_{\,\nu}=\fD_{\,\nu}+\fL_{\,\nu}
\, , \label{3.1}\end{equation}
in which the first term is a ``chemical'' contribution, representing the  
effect of any neutron superfluid particle creation or destruction, which 
is  given by 
\begin{equation}\fD{_{\!\nu}} =\muN_{\!\nu}\nabla_{\!\mu} \nN^\mu
\, .\label{3.2}\end{equation}
The last term in (\ref{3.1}) is a ``mechanical'' contribution, allowing 
for drag or pinning forces exerted on the vortices by the crust and 
balanced by the Magnus effect, according to the formula  
\begin{equation}\fL{_{\!\nu}}=\nN{^\mu}\wN_{\,\mu\nu}
\, ,\label{3.3}\end{equation}
using the notation 
\begin{equation}\wN_{\,\mu\nu}=2\nabla_{\![\mu}\muN_{\,\nu]}
\label{2.22}\end{equation}
for the vorticity 2-form of the superfluid neutrons. It is to be 
noted that this is not the mesoscopic (intervortex) superfluid vorticity, 
which simply vanishes, but the average vorticity on a macroscopic scale  
that is large compared with the spacing (typically a very small fraction  
of a cm.) between the superfluid vortices. For a very accurate 
treatment it would be necessary to take account of the macroscopic 
anisotropy resulting from the effective tension of these vortices, 
as has already been done~\cite{CarterLanglois95b,Carter00a} for the 
case a single constituent, but for the discussion of global evolution 
on timescales long compared with the stellar oscillation periods 
(a small fraction of a second) such an effect seems unlikely to be 
important. 
 
Although the complete expression (\ref{2.20}) is not so simple, it is 
to be observed that the time component in the ``normal'' rest frame 
(representing the rate of working on the ``normal'' constituent) as 
obtained by contraction with the relevant unit vector $\uC{^\nu}$ has 
the comparitively simple form 
\begin{equation}\uC{^\nu}\fC{_\nu}=\uC{^\nu}\muC{_\nu}\nabla_{\!\mu}
\nC{^\mu}-\Theta \nabla_{\!\mu} s^\mu
\, .\label{3.0}\end{equation}
 
If we were to impose the variation principle to the effect that the 
system should be invariant with respect to arbitrary worldline  
displacements (as specified by the independent fields $\xiC^{\,\nu}$  
and $\xiN^{\,\nu}$) it would follow that each of the forces   
$\fC_{\,\mu}$ and $\fN_{\,\nu}$ would have to vanish. However it is 
evident from the identity (\ref{2.23}) that we cannot adopt such a 
restrictive postulate in a model designed to treat the effect of  
pulsar slowdown due to a torque attributable to coupling to an  
external electromagnetic field that is removing angular momentum by  
radiation to infinity. As well as the intrinsically non-conservative  
magnetic torque contribution to $\fE_{\,\nu}$ it is also  
important~\cite{CLS00} to include a contribution to allow for the 
effect of the elastic solidity in the crust, which  is not 
incorporated into the simple fluid type model included here (and 
which would require the use of a much more elaborate model~\cite{C89} 
for its detailed evaluation).  
 
Although our ultimate purpose is to allow for a non vanishing external 
torque force, whatever force law we assume must be such that if the 
external force $\fE_{\,\nu}$ were somehow switched off so as to leave an 
effectively isolated system, the second law of thermodynics (no decrease 
of entropy in an isolated system) would be respected, i.e. we must have 
$\fE_{\,\nu}=0\Rightarrow \nabla_{\!\nu}d^\nu\geq 0$. It can be seen that 
this is equivalent to the requirement of positivity of the right hand side 
of the identity 
\begin{equation}\Theta\nabla_{\!\mu} s^\mu +\uC{^\nu}\fE{_{\!\nu}} 
=\uC{^\nu}(\muN{_\nu} -\muC{_\nu})\nabla_{\!\mu}\nN{^\mu} 
 +\uC{^\nu}\fL{_{\!\nu}} \, ,\label{3.8} \end{equation}
that is obtained from (\ref{3.0}), taking account of the total baryon 
conservation law (\ref{2.3}). Since they involve very different physical 
processes, one comes to the conclusion that each of the two terms on the  
right of (\ref{3.8}) must satisfy its own separate positivity condition. 
 
The positivity requirement for the first of these terms is presumably 
to be attributed to a crust particle creation law of the form 
\begin{equation}\nabla_{\!\mu} \nN^\mu=\Xi \uC^\nu(\muN{_\nu}-\muC{_\nu})
\label{3.10}\end{equation}
for some positive coefficient $\Xi$. Such a law is an obviously natural 
generalisation of the kind of creation rate formula that is familiar 
in chemical physics. In the present context what is involved is conversion 
of protons to neutrons by weak interactions, and the situation is 
complicated by the consideration that as far as the large scale mechanics 
of the neutron star is concerned, the effective rate may depend not just 
on microscopic processes, but also, when subduction is involved, on the 
rather messy process whereby the crust is broken up before it ultimately 
dissolves. 
 
To complete the specification of the system, all that remains is to find 
the appropriate ansatz for the mechanical force $\fL{_{\!\nu}}$. This 
problem is more delicate than that of the (effectively scalar) chemical 
case, since as well as the ``second law'' requirement 
$\uC^{\,\nu}\fL{_{\!\nu}}\geq 0$, the answer must respect the nature 
of the macroscopic vorticity 2-form $\wN_{\,\mu\nu}$ which 
although non vanishing (unlike the mesoscopic vorticity between vortices) 
cannot be arbitrary (as in an ordinary viscous fluid): to be consistent 
with the underlying superfluid nature of the neutron constituent, it 
must satisfy an algebraic degeneracy condition of the form  (\ref{26}) in 
order to be compatible with the existence of a well defined congruence 
of orthogonal 2-surfaces generated by (non vanishing) tangent vectors, 
$v^\nu$ say, such that $\wN_{\,\mu\nu}v^\nu=0$. It can be seen from the 
form of the defining relation (\ref{3.3}) that the obvious way to obtain 
this  degeneracy property is to take the force law to have the form 
$\fL{_{\!\nu}}=\wN_{\nu\sigma}v^\sigma$ for some suitably chosen vector 
$v^\nu$ which, to satisfy the ``second law'' requirement must satisfy 
$\uC^{\,\nu}\wN_{\,\nu\sigma}v^\sigma\geq 0$. The required ansatz 
can thus be taken to be given by $v_\mu=\alpha\wN_{\,\mu\nu}\uC^{\,\nu}$ 
for some positive coefficient $\alpha$. This result is conveniently 
expressible in terms of the rank-2 tensor $\perp^{\!\mu}_{\,\nu}$ 
of orthogonal projection with respect to the vortex 2-surface, 
which is given by 
\begin{equation}\perp^{\!\mu}_{\,\nu}=2(\wN{^{\rho\sigma}}
\wN_{\,\rho\sigma})^{-1} \wN{^{\lambda\mu}}\wN_{\,\lambda\nu} 
\, .\label{3.20}\end{equation}
We end up with an expression taking the form 
\begin{equation} \fL{_{\!\nu}}=\eta_{\rm r}\perp_{\nu\sigma}
\uC^{\,\sigma}\, ,\label{3.21}\end{equation}
for a positive resistive drag coefficient  $\eta_{\rm r}$ (given 
in terms of the previous coefficient $\alpha$ by $2\eta_{\rm r}$ 
$=\alpha \wN{^{\rho\sigma}}\wN_{\,\rho\sigma}$).

The generic class of dissipative models characterised by finite values of 
$\Xi$ and $\eta_{\rm r}$ has four different kinds of non dissipative 
limit. In the low reactivity limit $\Xi\rightarrow 0$ we have the 
non-transfusive limit characterised by the separate superfluid particle 
conservation law 
\begin{equation}\nabla_{\!\nu}\nN^{\,\nu}=0
\, ,\label{3.22}\end{equation}
whereas in the opposite high reactivity limit $\Xi\rightarrow\infty$ 
we have the chemical equilibrium limit characterised by 
\begin{equation}\uC^{\,\nu}\big(\muC_{\,\nu}-\muN_{\,\nu}\big)=0
\, .\label{3.23}\end{equation}
which is what would be expected in cases for which the (continental drift  
like) crust circulation responsible for the transfusion is characterised 
by timecales that are very long compared with those~\cite{Haensel92} of 
the  relevant weak (direct or inverse beta decay) interactions. For each 
of these conceivably relevant possibilities, we have the drag free limit  
$\eta_{\rm r}\rightarrow 0$  characterised by the condition of 
vanishing Magnus force, 
\begin{equation}\nN^{\mu}\wN_{\,\mu\nu}=0
\, ,\label{3.25}\end{equation}
or at the opposite extreme the perfect vortex pinning limit, 
$\eta_{\rm r}\rightarrow\infty$ characterised by the condition 
that the vortex worldsheets should be at rest with respect to the 
``normal'' (crust) background, 
\begin{equation}   \uC^{\mu}\wN_{\,\mu\nu}=0
\, .\label{3.26}\end{equation}
The actual evaluation of the drag coefficient $\eta_{\rm r}$, and the 
question of whether one or other of these simple extreme limits is realistic 
depends on delicate technical issues~\cite{Jones92,LinkEpsteinBaym93} 
whose definitive resolution is not entirely clear. A scenario of the type 
envisaged by Ruderman~\cite{R91}, as represented by Figure 1, and
described by (\ref{040}) (which seems to be appropriate for Vela) is what 
would be obtained in the case characterised by (\ref{3.26}), whereas the 
more recently proposed alternative scenario~\cite{CLS00} represented by 
Figure 2, and described by (\ref{041}), is what would be obtained in 
the case characterised by (\ref{3.25}).  
 
It is of course to be expected that such extreme scenarios will turn  
out in practise to be oversimplifications of a more complicated reality, 
whose description is likely to require modelisation with not just 
two~\cite{AC00} but many independently rotating  components, to allow 
for the variation of the chemical and mechanical coefficients $\Xi$ and 
$\eta_{\rm r}$ over a wide range of finite values as a function of depth 
in the star.  
 
\bigskip 
{\bf Acknowledgements} 
The author wishes to thank Silvano Bonazzola, David Langlois,  
Eric Gourghoulon, Pawel Haensel, Isaac Khalatnikov,  Reinhardt Prix,  
and David Sedrakian for conversations and collaboration. 
\medskip 
 
\vfill\eject

\end{document}